\documentclass[11pt]{article}
\usepackage{amsmath}
\usepackage{a4}
\usepackage{graphicx,psfrag,epsf}
\usepackage{psfrag}
\usepackage{algorithm}
\usepackage{marvosym}
\usepackage[noend]{algorithmic}
\usepackage{upgreek}
\usepackage{wrapfig}
\usepackage{boxedminipage}
\usepackage{pifont}
\usepackage{euscript}
\usepackage{dcolumn}
\usepackage{float}
\usepackage{fancybox}
\usepackage{rotating}
\usepackage{multirow}
\usepackage{latexsym}
\usepackage{amssymb}
\usepackage{txfonts}
\usepackage{amsmath}
\usepackage{epsfig}
\usepackage{color}
\usepackage{verbatim}
\usepackage{subfigure}
\usepackage{mathrsfs}
\usepackage{url}
\usepackage{lscape}
\usepackage{graphicx}
\usepackage{natbib}

\algsetup{indent=2em}
\usepackage{setspace}
\newtheorem{example}{Example}[section]

\newtheorem{alg}{Algorithm}[section]

\newtheorem{remark}{Remark}[section]
\newtheorem{lemma}{Lemma}[section]
\newtheorem{theorem}{Theorem}[section]
\newtheorem{proposition}{Proposition}[section]
\newtheorem{corollary}{Corollary}[section]

\newcommand{\ealg}{\end{alg}}
\newcommand{\balg}{\begin{alg}}

\newcommand{\bigO}{\EuScript{O}}

\newcommand{\bl}{\mathbf{l}}
\newcommand{\bs}{\mathbf{s}}

\newcommand{\bz}{\mathbf{z}}

\newcommand{\bX}{\mathbf{X}}
\newcommand{\bx}{\mathbf{x}}

\newcommand{\bq}{\mathbf{q}}

\newcommand{\by}{\mathbf{y}}
\newcommand{\bZ}{\mathbf{Z}}
\newcommand{\bu}{\mathbf{u}}

\newcommand{\ben}{\begin{enumerate}}
\newcommand{\een}{\end{enumerate}}
\newcommand{\beq}{\begin{equation}}
\newcommand{\eeq}{\end{equation}}
\newcommand{\ei}{\end{itemize}}
\newcommand{\bex}{\begin{example}}
\newcommand{\eex}{\end{example}}
\newcommand{\berem}{\begin{remark}}
\newcommand{\erem}{\end{remark}}
\newcommand{\beprop}{\begin{proposition}}
\newcommand{\eprop}{\end{proposition}}

\newcommand{\Var}{\text{Var}}

\renewcommand{\epsilon}{\varepsilon}
\renewcommand{\rho}{\varrho}
\renewcommand{\log}{\ln}
\renewcommand{\hat}{\widehat}
\renewcommand{\leq}{\leqslant}
\renewcommand{\geq}{\geqslant}

\newcommand{\argmax}{\mathop{\rm argmax}}
\newcommand{\argmin}{\mathop{\rm argmin}}

\newcommand{\U}{{\sf U}}

\newcommand{\Nor}{{\sf N}}

\newcommand{\Em}{\mathbb E}
\newcommand{\Pm}{\mathbb P}

\newcommand{\gvn}{\,|\,}

\newcommand{\ds}{\displaystyle}

\newcommand{\di}{\text{d}}
\newcommand{\tr}{\text{tr}}

\def\acro#1#2{\vskip4pt\hbox to\textwidth{\normalsize
\hbox to5pc{#1\hfill}\vtop{\advance\hsize by
-5pc\raggedright\noindent#2}}}

\def\symbol#1#2{\vskip4pt\hbox to\textwidth{\normalsize
\hbox to5pc{#1\hfill}\vtop{\advance\hsize by
-5pc\raggedright\noindent#2}}}

\newcommand{\I}{\,\mathbb{I}}

\newcommand{\iidsim}{\stackrel{\text{iid}}{\sim}}
\newcommand{\simiid}{\iidsim}

\newcommand{\diag}{\text{diag}}

\newcommand{\chk}[1]{}

\newcommand{\idef}{\stackrel{\text{def}}{=}}

\newcommand{\rank}{\text{rank}}

\newcommand{\bmu}{\boldsymbol\mu}
\newcommand{\bnu}{\boldsymbol\nu}
\newcommand{\bb}{\boldsymbol\beta}

\newcommand{\bpi}{\boldsymbol\pi}
\newcommand{\blam}{\boldsymbol\lambda}
\newcommand{\sJ}{\mathscr{J}}
\newcommand{\bPsi}{\boldsymbol\Psi}
\newcommand{\bfeta}{\boldsymbol\eta}

\title{The Normal Law Under Linear Restrictions: Simulation and Estimation via Minimax Tilting}
\author{Z. I. Botev\\
{\small The University of New South Wales,  \texttt{botev@unsw.edu.au}}}
\date{}

\begin{document}
\maketitle
\begin{abstract}

  Simulation from the truncated multivariate normal distribution in high dimensions is a recurrent problem  in statistical computing, and is typically  only feasible using approximate MCMC sampling. In this article	we propose a  minimax  tilting method for exact iid simulation from the truncated multivariate normal distribution. The new methodology provides both  a method for simulation  and an efficient estimator to hitherto intractable Gaussian integrals. We prove that the estimator possesses a rare  vanishing relative error asymptotic  property. 
Numerical experiments suggest that the proposed scheme is accurate in a wide range of setups for which competing estimation schemes fail.  
	We  give an application to exact iid  simulation from the Bayesian  posterior of the probit regression model.

\end{abstract}

\section{Introduction}
More than a century ago  Francis \citet{galton1889natural} observed that he scarcely knows ``anything so apt to impress the imagination as the wonderful form of cosmic order expressed by the law of frequency of error.  The law would have been personified by the  Greeks if they had known of it."   

 In this article we address some hitherto intractable computational problems related  to the $d$-dimensional multivariate normal law under linear restrictions:
\begin{equation}
\label{prob}\textstyle
f(\bz)=\frac{1}{\ell}\;\exp\left(-\frac{1}{2} \bz^\top\bz\right) \I\{\bl\leq A\bz\leq \bu\},\quad \bz=(z_1,\ldots,z_d)^\top,\quad A\in \mathbb{R}^{m\times d},\quad \bu,\bl\in\mathbb{R}^m\;,
\end{equation}
where $\I\{\cdot\}$ is the indicator function, $\rank(A)=m\leq d$, and  $\ell=\Pm(\bl\leq A\bZ\leq \bu)$ is the probability that  a random vector $\bZ$ with standard normal distribution in $d$-dimensions (that is, $\bZ\sim \Nor(\mathbf{0},I_d)$) falls in the $\mathsf{H}$-polytope defined by the linear inequalities. 

 Aesthetic considerations aside, the problem of estimating $\ell$ or simulating from $f(\bz)$  arises frequently in various contexts such as:  Markov random fields \citep{RSSB:RSSB12055}; inference for spacial processes \citep{wadsworth2014efficient}; likelihood estimation for max-stable processes  \citep{huser2013composite,genton2011likelihood};
 computation of simultaneous confidence bands \citep{azais2010simultaneous};
uncertainty regions for latent Gaussian models \citep{RSSB:RSSB12055}; fitting mixed effects models with censored data  \citep{RSSC:RSSC1007}; and probit regression \citep{albert1993bayesian},   to name a few. 


For the reasons outlined above, the problem of estimating $\ell$ accurately has received considerable attention. For example,  \cite{RSSB:RSSB625,miwa2003evaluation,gassmann2003multivariate,genz2004numerical,hayter2012evaluation,hayter2013evaluation} and \cite{nomura2014evaluation} consider
approximation methods for  special cases (orthant, bivariate, or trivariate probabilities) and  
\cite{geweke1991efficient,genz1992numerical,joe1995approximations,vijverberg1997monte,sandor2004alternative,nomura2012computation} consider estimation schemes applicable for general $\ell$. 
 Extensive comparisons amongst the numerous proposals in the literature
\citep{genz2009computation,gassmann2002computing,genz2002comparison} indicate the method of \cite{genz1992numerical} is the most accurate across a wide range of test problems of medium and large dimensions. Even in low dimensions ($d\leq 7$),  the method compares favorably with highly specialized routines for orthant probabilities \citep{miwa2003evaluation,RSSB:RSSB625}. 
For this reason, Genz' method is the default choice  across different software platforms like \textsf{Fortran}, \textsc{Matlab}$^{\circledR}$ and \textbf{\textsf{R}}. 

One of the goals of this article is to propose a new  methodology, which 
not only yields an unbiased estimator orders of magnitude less variable 
than the Genz estimator, but also works reliably in cases where the Genz 
estimator and other alternatives fail to deliver meaningful estimates 
(e.g., relative error close to 100\%)
\footnote{ \textsc{Matlab}$^{\circledR}$ and \textbf{\textsf{R}}  
implementations  are available
from \textsc{Matlab$^{\circledR}$ Central}, {\scriptsize \url{http://
www.mathworks.com/matlabcentral/fileexchange/53796}}, and the CRAN repository (under the name {\scriptsize\texttt{TruncatedNormal}}), as well 
as from the author's website: {\scriptsize\url{http://web.maths.unsw.edu.au
/~zdravkobotev/}}}. \enlargethispage{1cm}


The obverse to the problem of estimating $\ell$ is simulation from the truncated multivariate normal $f(\bz)$. Despite the close relation between the two problems, they have rarely been studied  concurrently \citep{botts2013accept,chopin2011fast,fernandez2007perfectly,philippe2003perfect}. Thus, another  goal of this article is to provide an exact accept-reject sampling scheme for simulation from $f(\bz)$ in high dimensions, which traditionally calls for approximate MCMC simulation. Such a scheme can either  obviate the need for 
Gibbs sampling  \citep{fernandez2007perfectly}, or can be used to accelerate Gibbs sampling  through the blocking of hundreds of highly dependent variables \citep{chopin2011fast}. Unlike existing algorithms,
the accept-reject sampler proposed in this article enjoys high acceptance rates in over one hundred  dimensions, and takes about the same time as one cycle of  Gibbs  sampling.

 The gist of the method is to find an exponential tilting of a suitable importance sampling measure by solving a minimax (saddle-point) optimization problem. 
The  optimization   can be solved efficiently, because it exploits \mbox{log-concavity} properties of the normal distribution.  The method permits us to construct an estimator with a tight 
deterministic bound on its relative error and a concomitant exact stochastic confidence interval.
Our  importance sampling proposal builds on the celebrated Genz construction, but the addition of the minimax tilting ensures that the new estimator  enjoys 
theoretically better   variance properties than the Genz estimator. In an appropriate asymptotic 
tail  regime, the minimax  tilting yields an 
estimator with vanishing relative error (VRE)   property \citep{kroese2011handbook}. 
 Within the light-tailed exponential family, Monte Carlo estimators rarely possess the valuable VRE property \citep{l2010asymptotic} and as yet  no estimator of $\ell$ with such properties has been proposed. 
  The VRE property   implies, for example, that the new accept-reject instrumental density converges in total variation to the target density $f(\bz)$, rendering sampling  in the tails of the truncated normal distribution asymptotically feasible.  In this article we focus on the multivariate normal law due to its central position in statistics, but the proposed methodology can be easily generalized to other multivariate elliptic distributions.

\section{Background on Separation of Variables Estimator}
\label{sec:SOV}
We first briefly describe the separation of variables (SOV) estimator of \cite{genz1992numerical} (see also \cite{geweke1991efficient}).  Let 
$
A=LQ^\top
$
be the LQ decomposition of the matrix $A$, 
where $L$ is $m\times d$ lower triangular with nonnegative entries down the main diagonal and $Q^\top=Q^{-1}$ is $d\times d$ orthonormal. A simple change of variable  $\bx\leftarrow Q^\top\bz$ then yields:
\[
\ell=\Pm(\bl\leq L\bZ\leq \bu)=\int_{\mathbf{l}\leq L\bx\leq \bu}  \phi(\bx;\mathbf{0},I)\,\di \bx,
\]
where $\phi(\bx;\bmu,\Sigma)$ denotes the pdf of the $\Nor(\bmu,\Sigma)$ distribution.
For simplicity of notation, we henceforth assume that $m=d$ so that $L$ is full rank. The case of $m<d$ is considered later in the experimental section. 
\cite{genz1992numerical} decomposes the region $\mathscr{C}=\{\bx: \mathbf{l}\leq L\bx\leq \bu\}$  sequentially as follows:
\begin{align*}
 \tilde l_1\idef \frac{l_1}{L_{11}}\leq x_1& \leq \frac{u_1}{L_{11}}\idef \tilde u_1 \\
 \tilde l_2(x_1)\idef\frac{l_2-L_{21}x_1}{L_{22}}\leq x_2& \leq \frac{u_2-L_{21}x_1}{L_{22}}\idef \tilde u_2(x_1) \\
\vdots&\\
\tilde l_d(x_1,\ldots,x_{d-1})\idef\frac{l_d-\sum_{j=1}^{d-1}L_{dj}x_j}{L_{dd}}\leq x_d& \leq \frac{u_d-\sum_{j=1}^{d-1}L_{dj}x_j}{L_{dd}}\idef \tilde u_d(x_1,\ldots,x_{d-1}) \\
\end{align*}
This decomposition  motivates the separation of variables estimator of $\ell$
\begin{equation}
\label{est}
\hat{\ell}=\frac{ \phi(\bX;\mathbf{0},I)}{g(\bX)},\qquad \bX\sim g(\bx)
\end{equation} 
where $g$ is an importance sampling density over the set $\mathscr{C}$ and in the SOV form
\begin{equation}
g(\bx)=g_1(x_1)g_2(x_2\gvn x_1)\cdots g_d(x_d\gvn x_1,\ldots,x_{d-1}),\qquad \bx\in\mathscr{C} .
\end{equation}
We denote the measure corresponding to $g$ by $\Pm_\mathbf{0}$.
The Genz SOV estimator, which we denote by $\mathring{\ell}$ to distinguish it from the more general $\hat \ell$, is obtained by selecting for all $k=1,\ldots,d$
\begin{equation}
\label{genz choice}
g_k(x_k\gvn x_1,\ldots,x_{k-1})\varpropto \phi(x_k;0,1)\times \I\{\tilde l_k\leq x_k\leq \tilde u_k\}
\end{equation}
Denoting by $\Phi(\cdot)$  the cdf of the standard normal distribution, this gives the following. 
\begin{alg}[SOV estimator]~
\label{SOV}
\begin{algorithmic}
\REQUIRE {The lower triangular $L$ such that $A=LQ^\top$, bounds $\mathbf{l}$, $\bf{u}$, and uniform sequence $U_1,\ldots,U_{d-1}\simiid\U(0,1)$. }

\FOR {$k=1,2,  \ldots, d-1$}
\STATE{Simulate 
$X_k\sim\Nor(0,1)$ conditional on 
$
\tilde l_k(X_1,\ldots,X_{k-1})\leq X_k\leq \tilde u_k(X_1,\ldots,X_{k-1})
$ using the inverse transform method. That is, set 
\[
X_k=\Phi^{-1}\left( \Phi(\tilde l_k)+U_k\left(\Phi(\tilde u_k)-\Phi(\tilde l_k)\right)  \right).
\]
}
\ENDFOR
\RETURN{$\ds
\mathring{\ell}=\prod_{k=1}^d \left[\Phi( \tilde u_k(X_1,\ldots,X_{k-1}))-\Phi( \tilde l_k(X_1,\ldots,X_{k-1}))\right]
$.}
\end{algorithmic}
\end{alg}
The algorithm can be repeated $n$  times to obtain 
the iid sample $\mathring{\ell}_1,\ldots,\mathring{
\ell}_n$ used for the construction of  the unbiased point estimator 
$\bar{\ell}=(\mathring{\ell}_1+\cdots+\mathring{\ell}_n)/n$ 
and its approximate 95\% confidence interval $(\bar{\ell}\pm 1
.96 \times S/\sqrt{n})$, where $S$ is the sample standard 
deviation of $\mathring{\ell}_1,\ldots,\mathring{
\ell}_n$. 


\subsection{Variance Reduction via Variable Reordering}
\label{ordering}
 \cite{genz2009computation} suggest the following  improvement of the SOV algorithm. 
Let $\boldsymbol{\pi}=(\pi_1,\ldots,\pi_d)$ be a permutation of the integers $1,\ldots,d$ and denote the corresponding permutation matrix $P$ so that  $P(1,\ldots,d)^\top=\bpi$. It is clear that for any $\bpi$
we have 
$
\ell=\Pm(P\bl\leq PA\bZ \leq P\bu)
$. Hence, to estimate $\ell$, one can input in  the SOV Algorithm~\ref{SOV} the permuted bounds and matrix: $\bl\leftarrow P\bl, \bu\leftarrow P\bu$, and $A\leftarrow PA$. This results in an unbiased estimator $\mathring{\ell}(\bpi)$ whose variance  will depend on $\bpi$ --- the order in which this high-dimensional  integration is carried out. Thus, we would like to choose the $\bpi^*$ amongst all possible permutations so that
\[
\bpi^*=\argmin_{\bpi}\Var(\mathring{\ell}(\bpi))
\]
This is an intractable  combinatorial optimization problem whose objective function is not even available. 
Nevertheless,  \cite{genz2009computation} propose a heuristic for finding an acceptable approximation to $\bpi^*$. We henceforth assume that this variable reordering heuristic is always applied as a preprocessing step to the SOV Algorithm~\ref{SOV} so that  the matrix $A$ and the bounds $\bl$ and $\bu$ are already in permuted form. We will revisit variable reordering in the numerical experiments  in  Section~\ref{numerics}. 


 The main limitation  of the estimator $\mathring{\ell}$ (with or without variable reordering) is 
that $\Var(\mathring{\ell})$ is unknown and its estimate $S^2$
 can be notoriously unreliable in the sense that the observed 
$S^2$ may be very small, while the true $\Var(\mathring{\ell})
$ is huge \citep{kroese2011handbook,botev2013markov}. Such  examples for which  $\mathring{\ell}$ fails to deliver meaningful estimates of $\ell$ will be given in the numerical Section~\ref{numerics}. 

\subsection{Accept-Reject Simulation}

The SOV approach described above 
suggests that we could   simulate  from $f(\bz)$ exactly by using $g(\bx)$ as an instrumental density in the following accept-reject scheme  \citep[Chapter 3]{kroese2011handbook}.

\begin{alg}[Accept-Reject Simulation from $f$]~
\label{accept-reject}
\begin{algorithmic}
\REQUIRE {Supremum of likelihood ratio 
$
c=\sup_{\bx\in\mathscr{C}} \phi(\bx;\mathbf{0},I)/g(\bx). 
$
}
\STATE{Simulate $U\sim\U(0,1)$ and $\bX\sim g(\bx)$, independently.}
\WHILE{$c U>\phi(\bX;\mathbf{0},I)/g(\bX)$}
\STATE{Simulate $U\sim\U(0,1)$ and $\bX\sim g(\bx)$, independently.}
\ENDWHILE
\RETURN{$\bX$, an outcome from the truncated multivariate normal density $f$ in \eqref{prob}.}
\end{algorithmic}
\end{alg}
Of course, the accept-reject scheme will only be usable if the probability of acceptance $\Pm_\mathbf{0}(c U\leq\phi(\bX;\mathbf{0},I)/g(\bX))=\ell/c$ is high and simulation from $g$ is fast. 
Thus, this scheme  presents two significant challenges which need resolution. The first one is the computation of the constant $c$ (or a very tight upper bound of it) in finite time. Locating the global maximum  of the likelihood ratio  $\phi(\bx;\mathbf{0},I)/g(\bx)$ may be an intractable problem --- a local maximum will yield an incorrect sampling scheme. The second challenge is to select  an instrumental  $g$ so that the acceptance probability is not
prohibitively small (a ``rare-event" probability). Unfortunately,  the obvious choice \eqref{genz choice}  resolves neither of these challenges \citep{hajivassiliou1998}.
Other  accept-reject schemes \citep{chopin2011fast}, while excellent in one and two dimensions, ultimately have acceptance rates of the order $\mathscr{O}(2^{1-d})$ rendering them unusable for this type of problem with, say, $d=100$. 
 We now address  these issues concurrently in the next section.

\section{Minimax Tilting}
Exponential tilting is a prominent technique in simulation \citep{l2010asymptotic,kroese2011handbook}. For a given light-tailed  probability density $h(y)$  on $\mathbb{R}$,  we can associate with $h$ its   exponentially tilted version $h_\mu(y)=\exp\left(\mu y -K(\mu)\right) h(y)$, where  $K(\mu)=\log\Em\exp(\mu X)<\infty$, for some $\mu$ in an open set, is the  cumulant generating function. 
%
%
%
%
%
%
For example, the exponentially tilted version of $\phi(\bx;\mathbf{0},I)$  is $\exp\left(\bmu^\top\bx -K(\bmu)\right)\phi(\bx;\mathbf{0},I)=\phi(\bx;\bmu,I)$.
 Similarly, the tilted version of   \eqref{genz choice} yields
\begin{equation} 
\label{tilted choice}
 g_k(x_k;\mu_k\gvn x_1,\ldots,x_{k-1})=\frac{\phi(x_k;\mu_k,1)\times\I\{\tilde l_k\leq x_k\leq \tilde u_k\}}{\Phi( \tilde u_k-\mu_k)-\Phi( \tilde l_k-\mu_k)}
\end{equation}
To simplify the notation in the subsequent analysis,  let 
\begin{align}
\label{psi}
 \psi(\bx;\bmu)&\idef-\bx^\top\bmu+\frac{\|\bmu\|^2}{2}+\sum_k\log \left(\Phi( \tilde u_k(x_1,\ldots,x_{k-1})-\mu_k)-\Phi( \tilde l_k(x_1,\ldots,x_{k-1})-\mu_k)\right)
\end{align}
Then, the tilted version of estimator \eqref{est} can be written as
$\hat\ell=\exp\left(\psi(\bX;\bmu)\right)$ with $\bX\sim  \Pm_{\bmu}$,
 where $\Pm_{\bmu}$ is the measure with pdf
 $
g(\bx;\bmu) \idef\prod_{k=1}^d g_k(x_k;\mu_k\gvn x_1,\ldots,x_{k-1}).
$
 It is  now clear that the statistical properties of $\hat\ell$ depend on the tilting parameter $\bmu$.
 There is a large  literature  on the best way to select the tilting parameter $\bmu$; see  \cite{l2010asymptotic} and the references therein. A recurrent theme in all works is the efficiency of the estimator $\hat\ell$ in a tail asymptotic regime where $\ell\downarrow 0$ is a rare-event probability --- precisely the setting that makes current accept-reject schemes inefficient.  Thus, before we continue,  we briefly recall  the three widely used criteria for assessing  efficiency in estimating tail probabilities.

 The weakest type of efficiency and the most commonly encountered in the design of importance sampling schemes \citep{kroese2011handbook} is  logarithmic efficiency. The estimator $\hat\ell$ is said to be \emph{logarithmically or weakly efficient} if
\[
\liminf_{\ell\downarrow 0}\frac{\log \Var(\hat\ell)}{\log\ell^2}\geq 1
\]
 The second and stronger type of efficiency is  \emph{bounded relative error}, 
 \[
\limsup_{\ell\downarrow 0} \frac{\Var(\hat\ell)}{\hat\ell^2} \leq \mathrm{const.}<\infty.
\]
Finally, the  best one can hope for in an asymptotic regime is 
the highly desirable \emph{vanishing relative error} (VRE) property:
\begin{equation*}
\limsup_{\ell\downarrow 0} \frac{\Var(\hat\ell)}{\hat\ell^2}=0\;.
\end{equation*}
An estimator is \emph{strongly efficient} if it exhibits either  bounded relative error or VRE. In order to achieve one of these efficiency criteria,
most  methods 
\citep{l2010asymptotic} rely on the derivation of an 
analytical asymptotic approximation to the relative error $\Var(\hat\ell)/\ell^2$,  whose  behavior is then controlled using the tilting parameter. The strongest type of efficiency VRE is uncommon for light-tailed probabilities, and is typically only achieved within a state-dependent importance sampling framework \citep{l2010asymptotic}.

  Here we take a different  tack, one that exploits features unique to the problem at hand and that will yield efficiency gains in both an asymptotic and non-asymptotic regime.  A key result in this direction is  the following Lemma~\ref{lemma}, whose proof is given in the appendix.
\begin{lemma}[Minimax Tilting]
\label{lemma}
The optimization program
\[
\inf_{\bmu}\sup_{\bx\in \mathscr{C}}\psi(\bx;\bmu)
\]
is a saddle-point problem with a unique solution given by the concave optimization program:
\begin{equation}
\label{optim}
\begin{split}
 (\bx^*,\bmu^*)=&\argmax_{\bx,\bmu} \psi(\bx;\bmu)\\
\textrm{subject to: }&  \frac{\partial \psi}{\partial\bmu} =\mathbf{0},\quad\; \bx \in \mathscr{C}
\end{split}
\end{equation}
\end{lemma}
Note that \eqref{optim}  minimizes with respect to $\bmu$ the worst-case behavior of the likelihood ratio, namely $\sup_{\bx\in \mathscr{C}}\exp\left(\psi(\bx;\bmu)\right)$. The  lemma states we can both easily locate  the global worst-case behavior  of the likelihood ratio, and simultaneously locate (in finite computing time) the global minimum with respect to $\bmu$. Prior to analyzing the theoretical properties of minimax tilting, we first explain how to implement the minimax method in practice.

\paragraph{Practical Implementation.}
How do we find the solution of  \eqref{optim} numerically? Without the constraint $\bx\in\mathscr{C}$, the solution to \eqref{optim} would be obtained by
solving the nonlinear system of equations $\nabla\psi(\bx;\bmu)=\mathbf{0}$, where the gradient is with respect to the vector $(\bx,\bmu)$. To show why this is the case, we introduce the following notation. Let $D=\mathrm{diag}(L),\;\breve L=D^{-1}L$, and
\[
\Psi_j\idef \frac{\phi(\tilde l_j;\mu_j,1)-\phi(\tilde u_j;\mu_j,1)}{\Pm(\tilde l_j-\mu_j\leq Z\leq \tilde u_j-\mu_j)},
\]
\[
\Psi'_j\idef \frac{\partial\Psi_j}{\partial\mu_j}=\frac{(\tilde l_j-\mu_j)\phi(\tilde l_j;\mu_j,1)-(\tilde u_j-\mu_j)\phi(\tilde u_j;\mu_j,1)}{\Pm(\tilde l_j-\mu_j\leq Z\leq \tilde u_j-\mu_j)}-\Psi_j^2\;.
\]
Then,  the gradient equation $\nabla\psi(\bx;\bmu)=\mathbf{0}$ can be written as 
\begin{equation}
\label{nonlin}
\frac{\partial \psi}{\partial \bx}=-\bmu+(\breve L^\top-I)\bPsi=\mathbf{0},
\qquad  \frac{\partial \psi}{\partial \bmu}=\bmu-\bx+\bPsi=\mathbf{0}\;,
\end{equation}
and the  Jacobian matrix  has elements: 
\begin{equation}
\label{Jacobian}
\frac{\partial^2 \psi}{\partial \bmu^2}=I+\mathrm{diag}\left(\bPsi'\right),\quad\frac{\partial^2 \psi}{\partial \bmu\partial \bx}=(\breve L-I)\mathrm{diag}(\bPsi')-I,\quad 
\frac{\partial^2 \psi}{\partial \bx^2}=(\breve L-I)^\top\mathrm{diag}\left(\bPsi'\right) (\breve L-I)\;.
\end{equation}
The Karush-Kuhn-Tucker  equations give the necessary and sufficient condition for the global  solution $(\bx^*,\bmu^*)$ of \eqref{optim}:
\newcommand{\Eta}{\boldsymbol\eta}
\newcommand{\Tau}{\boldsymbol\tau}
\begin{equation}
\label{KKT}
\begin{split}
\partial \psi/\partial \bmu=\mathbf{0},\;\,\qquad \partial \psi/\partial \bx- \breve{L}^\top\Eta_1+\breve{L}^\top\Eta_2 &=\mathbf{0} \\
\Eta_1\geq\mathbf{0},\quad L\bx -\bu\leq \mathbf{0}, \quad   \Eta_1^\top(L\bx-\bu)&=\mathbf{0}\\
\Eta_2\geq \mathbf{0},\quad-L\bx+\bl\leq \mathbf{0},\quad  \Eta_2^\top(L\bx-\bl)&=\mathbf{0},
\end{split}
\end{equation}
where $\Eta_1,\Eta_2$ are Lagrange multipliers. 

Suppose we find the unique solution of the nonlinear system \eqref{nonlin} using, for example, a trust-region Dogleg method \citep{powell1970hybrid}. If we denote the solution to \eqref{nonlin} by $(\breve\bx,\breve\bmu)$, then  the Karush-Kuhn-Tucker equations imply that $(\breve\bx,\breve\bmu)=(\bx^*,\bmu^*)$ if and only if
$(\breve\bx,\breve\bmu)\in \mathscr{C}$ or equivalently $\Eta_1=\Eta_2=\mathbf{0}$.  
If, however, the solution $(\breve\bx,\breve\bmu)$ to \eqref{nonlin} does not lie in $\mathscr{C}$, then  $(\breve\bx;\breve\bmu)$ will be  suboptimal and, in order to compute $(\bx^*;\bmu^*)$,  one has to  use a  constrained convex optimization solver. 
This observation then leads to the following procedure.
\begin{alg}[Computation of optimal pair $(\bx^*,\bmu^*)$]~
\label{dog-leg}
\begin{algorithmic}
\STATE{Use Powell's (1970) Dogleg method on 
\eqref{nonlin} with Jacobian \eqref{Jacobian} to find $(\breve\bx,\breve\bmu)$.}
\IF{$(\breve\bx,\breve\bmu)\in\mathscr{C}$}
\STATE{$(\bx^*,\bmu^*)\leftarrow (\breve\bx,\breve\bmu)$}
\ELSE
\STATE{Use a  convex  solver to find $(\bx^*,\bmu^*)$, where $ (\breve\bx,\breve\bmu)$ is the initial guess.}
\ENDIF
\RETURN{$(\bx^*,\bmu^*)$}
\end{algorithmic}
\end{alg}
 Numerical experience suggests  almost always  $(\breve\bx,\breve\bmu)$ happens to lie in $\mathscr{C}$ and there is no need to do any additional computation over and above  Powell's (1970) trust-region method.

\section{Theoretical Properties of  Minimax Tilting}
There are a number of reasons why the minimax program \eqref{optim}  is an excellent way of selecting the tilting parameter. The first one shows that, unlike its competitors, the proposed estimator, 
\begin{equation}
\label{tilted est}
\hat\ell=\exp\left(\psi(\bX;\bmu^*)\right), \qquad \bX\sim \Pm_{\bmu^*},
\end{equation}
achieves the best possible  efficiency in a tail  asymptotic regime.

Let $\Sigma=AA^\top$ be a full rank covariance matrix. Consider the tail
 probability
$\ell(\gamma)=\Pm(\bX\geq \gamma\bl)$, where $\bX\sim\Nor(\mathbf{0},\Sigma)$ and $\gamma>0,\; \bl>\mathbf{0}$. We show that the estimator \eqref{tilted est} exhibits strong efficiency in estimating $\ell(\gamma)$ as $\gamma\uparrow\infty$. To this end, we first introduce the following simplifying notation.

\newcommand{\pl}{\mathbf{p}}
Similar to the variable reordering in Section~\ref{ordering},
suppose that $P$ is a permutation matrix which maps the vector $(1,\ldots,d)^\top$ into the  permutation
$\bpi=(\pi_1,\ldots,\pi_d)^\top$, that is, $P(1,\ldots,d)^\top=\bpi$.  
Let $L$ be the lower triangular factor of $P\Sigma P^\top=LL^\top$ and $\pl=P\bl$. It is clear that 
\[
\ell(\gamma)=\Pm(P\bX\geq\gamma P\bl)=\Pm(L\bZ\geq \gamma \pl )
\]
 for any permutation $\bpi$. For the time being, we leave 
$\bpi$ unspecified, because unlike in Section~\ref{ordering}, here we do not use $\bpi$ to minimize the variance of the estimator, but to simplify the notation in our efficiency analysis.

  Define the convex quadratic programming problem:
\begin{equation}
\label{qqp}
\begin{split}
\min_{\bx}\;\;& \frac{1}{2}\|\bx\|^2\\
\textrm{subject to: }& L\bx\geq \gamma\pl
\end{split}
\end{equation}
 The Karush-Kuhn-Tucker equations, which are a necessary and sufficient condition to find the solution of \eqref{qqp}, are given by:
 \begin{equation}
\label{kkt}
\begin{split}
 \bx-L^\top\blam&=\mathbf{0} \\
\blam\geq \mathbf{0},\;\;\gamma\pl-L\bx&\leq\mathbf{0}\\
\blam^\top(\gamma\pl-L\bx)&=0\;,
\end{split}
\end{equation}
where  $\blam\in\mathbb{R}^d$ is a Lagrange multiplier vector.
Suppose the number of active constraints in \eqref{qqp} is $d_1$ and the number of inactive constraints is $d_2$, where $d_1+d_2=d$. Note that since $L\bx\geq\gamma\pl>\mathbf{0}$, the number of active constraints $d_1\geq 1$, because otherwise $\bx=\mathbf{0}$ and $L\bx=\mathbf{0}$, reaching a contradiction.

Given the partition  $\blam=(\blam_1^\top,\blam_2^\top)^\top$ with $\dim(\blam_1)=d_1$ and $\dim(\blam_2)=d_2$,
 we now choose $\bpi$ such that all the active constraints in \eqref{kkt} correspond to $\blam_1>\mathbf{0}$  and all the 
inactive ones to $\blam_2=\mathbf{0}$. Similarly, we define a partitioning  for $\bx,\pl$, and the  lower triangular
\[
L=\left(\begin{array}{cc}
L_{11} & O\\
L_{21}& L_{22}
 \end{array}\right)\;.
\]

Note that the only reason   for introducing the above variable reordering via the permutation matrix $P$ and insisting that  all active constraints of \eqref{qqp} are collected 
in the upper part of vector $\blam$ is notational convenience and simplicity. At the cost of some generality,  this preliminary variable reordering  allows us to state and prove the efficiency result in the following Theorem~\ref{BRE} in its simplest and neatest form.

\begin{theorem}[Strong Efficiency of Minimax Estimator]
\label{BRE}
Consider the estimation of the   probability  
\[
\ell(\gamma)=\Pm(\bX\geq\gamma\bl)=\Pm(L\bZ\geq \gamma\pl)
\]
where  $\bX\sim\Nor(\mathbf{0},\Sigma),\;\bZ\sim\Nor(\mathbf{0},I)$; and $LL^\top=P\Sigma P^\top,\;\pl=P\bl>\mathbf{0}$ are the permuted versions of $\Sigma,\bl$ ensuring that the Lagrange multiplier vector $\blam$ in  \eqref{kkt} satisfies $\blam_1>\mathbf{0}$ and $\blam_2=\mathbf{0}$. Define
\[
\bq\idef L_{21}L_{11}^{-1}\pl_1-\pl_2
\]
 and  let $\sJ$ be the set of indices for which the components of the vector $\bq$ are zero, that is, 
\begin{equation}
\label{J}
\sJ\idef\{j:q_j=0,\; j=1,\ldots,d_2\}
\end{equation}
If $\sJ=\emptyset$,  then the minimax estimator \eqref{tilted est}
 is a vanishing relative error estimator:
\[\textstyle
\limsup_{\gamma\uparrow\infty}\frac{\Var_{\bmu^*}(\hat\ell(\gamma))}{\ell^2(\gamma)}=0\;.
\]
Alternatively, if $\sJ\not=\emptyset$, then
$\hat\ell$ is  a bounded relative error estimator: 
\[\textstyle
\limsup_{\gamma\uparrow\infty}\frac{\Var_{\bmu^*}(\hat\ell(\gamma))}{\ell^2(\gamma)}<\textrm{const.}<\infty.
\] 
\end{theorem}
The theorem suggests that,  unless the covariance matrix $\Sigma$ has a very special structure, the estimator enjoys VRE. This raises the question: Is there a simple setting that guarantees VRE for any full-rank covariance matrix under any preliminary variable reordering?


The next result shows that when $\bl$ can be represented as a weighted linear combination of the columns of the covariance matrix $\Sigma=AA^\top$, then we always have VRE. 

\begin{theorem}[Minimax Vanishing Relative Error]
\label{VRE}
Consider the estimation of the tail  probability $\ell(\gamma)=\Pm(\gamma\bl\leq A\bZ\leq \boldsymbol\infty)$, where  $\bl= \Sigma \bl^*$ for some positive  weight $\bl^*>\mathbf{0}$. 
Then, the minimax estimator \eqref{tilted est} 
is a vanishing relative error estimator.

In contrast,
under the additional assumption $L^\top\bl^*>\mathbf{0}$ (strong positive covariance), where $L$ is the lower triangular factor of  $\Sigma=LL^\top$,   the SOV estimator $\mathring{\ell}$ is a bounded relative error estimator; otherwise, it is   a  divergent one\footnote{The symbols $f(x)\simeq g(x)$, $f(x)=\bigO(g(x))$, and $f(x)=o(g(x))$, as $x\uparrow\infty$ and $g(x)\not=0$, stand for
$\lim_{x\uparrow \infty}f(x)/g(x)= 1$, $\limsup_{x\uparrow \infty} |f(x)/g(x)|<\infty$, and $\lim_{x\uparrow \infty} f(x)/g(x)=0$, respectively. }:
\[
\textstyle  \frac{\Var_\mathbf{0}(\exp(\psi(\bX;\mathbf{0}))) }{\ell^2(\gamma)}\simeq \begin{cases}\bigO(1), &\textrm{if }     L^\top\bl^*>\mathbf{0}\\
	 \exp(\bigO(\gamma^2)+\bigO(\log\gamma)+\bigO(1)), &\textrm{otherwise}
	\end{cases}\;.
\] 
\end{theorem}
Note that  the permutation matrix $P$ plays no role in the statement of Theorem~\ref{VRE} (we can assume $P=I$), and that we do not assume  $\bl>\mathbf{0}$, but only that $\bl=\Sigma\bl^*$ for some $\bl^*>\mathbf{0}$.

In light of Theorems~\ref{BRE} and \ref{VRE}, for the obverse problem of simulation from the truncated multivariate normal, we  obtain the following result.
\begin{corollary}[Asymptotically Efficient Simulation]
\label{corollary}
 Suppose that  the  instrumental density in the Accept-Reject Algorithm~\ref{accept-reject} for simulation from 
\[
f(\bz)\varpropto \phi(\bz;\mathbf{0},I)\times\I\{A\bz\geq \gamma \bl\},
\] 
is given by
$
g(\bx;\bmu^*)
$. Suppose further that,  either  $\bl>\mathbf{0}$ and the corresponding estimator \eqref{tilted est} enjoys VRE, or $\bl=\Sigma\bl^*$ for some $\bl^*>\mathbf{0}$.
Then, the measure $ \Pm_{\bmu^*}$ becomes indistinguishable from the target  $\Pm$:
\[
\sup_{\mathscr{A}}|\Pm(\bZ\in \mathscr{A})-\Pm_{\bmu^*}(\bZ\in \mathscr{A})|\rightarrow 0,\qquad \gamma\uparrow\infty.
\]
\end{corollary}

 A second reason that recommends our choice of tilting parameter  is  that $\exp\left(\psi(\bx^*;\bmu^*)\right)$ is a nontrivial deterministic upper bound to $\ell$, that is,
$
\ell\leq \exp(\psi(\bx^*;\bmu^*))
$. 

 As a result, unlike many existing  estimators \citep{vijverberg1997monte,genz1992numerical}, we can construct an exact (albeit conservative) confidence interval for $\ell$ as follows. Let  $\epsilon>0$ be the desired  width of the $1-\alpha$ confidence interval and $\ell_\mathrm{L}\leq \ell$ be a lower bound to $\ell$. Then,  by Hoeffding's inequality for  $\bar\ell=(\hat\ell_1+\cdots+\hat\ell_n)/n$ with 
\begin{equation}
\label{Hoeffding}
n(\epsilon)= \big\lceil -\log(\alpha/2)\times (\exp(\psi(\bx^*;\bmu^*))-\ell_\mathrm{L})^2/(2\epsilon^2) \big\rceil,
\end{equation}  we  obtain:
$
\Pm_{\bmu^*}(\bar\ell-\epsilon\leq \ell\leq\bar\ell+\epsilon )\geq 1-\alpha
$.

As is widely-known \citep{kroese2011handbook}, 
the main weakness of any importance sampling estimator  $\bar\ell$ of  $\ell$ is the risk of severe  underestimation of $\ell$.  Thus, plugging  $\bar\ell$  (or even more conservatively, plugging zero) in place of $\ell_\mathrm{L}$ in the formula for $n$ above will yield a robust confidence interval $(\bar\ell\pm \epsilon)$. For practitioners who are not satisfied with such a heuristic approach,  we provide the following  deterministic lower bound to $\ell$.
\begin{lemma}[Cross Entropy Lower Bound]
\label{lower} Define  the product measure $\underline \Pm$ with pdf 
 \[
\underline\phi(\bx)\varpropto \phi(\bx;\bnu,\mathrm{\diag}^2(\boldsymbol\sigma))\times \I\{\bl\leq \bx\leq \bu\}\;,
\]
where  $\bnu$  and $\boldsymbol\sigma=(\sigma_1,\ldots,\sigma_d)^\top$ are location and scale parameters, respectively. Define
\[
\ell_\mathrm{L}=\sup_{\bnu,\boldsymbol\sigma}\frac{\exp\left(- \frac{1}{2}\tr(\Sigma^{-1} \underline\Var(\bX))-\frac{1}{2} \underline\Em[\bX]^\top \Sigma^{-1} \underline\Em[\bX]-\underline\Em[\log {\underline\phi(\bX)]}\right)}{(2\pi)^{d/2}|\det(A)|} \;,
\]
where $\Sigma=AA^\top$.
 Then,  $
\ell_\mathrm{L}\leq \ell
$
 is a variational lower
bound to $\ell$. In addition, under the conditions of Theorem~\ref{VRE}, namely, $(\bl,\bu)=(\gamma\Sigma\bl^*,\boldsymbol\infty)$ , we have that
$\ell_\mathrm{L}\uparrow\ell(\gamma)$ and
\begin{equation}
\label{total var}
\sup_{\mathscr{A}}|\Pm(\bZ\in \mathscr{A})-\underline\Pm(A^{-1}\bZ\in \mathscr{A})|\downarrow 0,\qquad \gamma\uparrow\infty\;.
\end{equation}
\end{lemma}
 Since simulation from $\underline\Pm$ is straightforward, one may be tempted to consider using $\underline\Pm$ as an alternative importance measure to
$\Pm_{\bmu^*}$.  Unfortunately, despite the similarity of the results in Theorem~\ref{VRE} and Lemma~\ref{lower}, the pdf $\underline\phi$ is not amenable to an accept-reject scheme for exact sampling from $f$ and
as an importance sampling measure it does not yield VRE.  Thus, the sole use of Lemma~\ref{lower} is for constructing an exact confidence interval and lower bound to $\ell$ in the tails of the normal distribution.

Note that under the conditions of  Theorem~\ref{VRE}, the minimax estimator  enjoys the bounded normal approximation property \citep{tuffin1999bounded}. That is, if $\bar\ell$ and $S^2$ are the mean and sample variance of the iid $\hat\ell_1,\ldots,\hat\ell_n$, and $F_n(x)$
is the empirical cdf of  $T_n=\sqrt{n}(\bar\ell-\ell)/S$, then
 we have the  Berry--Ess\'een bound, uniformly in $\gamma$:
\[
\sup_{x\in\mathbb{R},\gamma>0}|F_n(x)-\Phi(x)|\leq \mathrm{const.}/\sqrt{n}
\]
This Berry--Ess\'een bound implies that the coverage error of the approximate $(1-\alpha)$ level  confidence interval $\bar\ell\pm z_{1-\alpha/2}\times S/\sqrt{n}$ remains of the order $\bigO(n^{-1/2})$, even as $\ell\downarrow0$. Thus, if a lower bound  $\ell_\mathrm{L}$ is not easily available,  one can still rely on the  confidence interval derived from the central limit theorem.

Finally, in addition to the strong efficiency properties of the estimator, another reason that recommends the minimax   estimator is
that it permits us to tackle intractable simulation and estimation problems  as illustrated in the next section.

\section{Numerical Examples and Applications}
\label{numerics}
We begin by considering a number of test cases used throughout   the literature
 \citep{fernandez2007perfectly,RSSB:RSSB625,miwa2003evaluation}. 
We are interested in  both the efficient simulation of the Gaussian vector  $\bX=A\bZ\sim \Nor(\mathbf{0},\Sigma)$ conditional on  $\bX\in \mathscr{A}$, and the estimation of $\ell$ in \eqref{prob}.

In all  examples we compare the separation-of-variables (SOV) estimator of Genz with the proposed minimax-exponentially-tilted (MET) estimator.   
We note that initially we considered a comparison with  other estimation schemes  such as the radially symmetric approach of \cite{nomura2012computation}  and the specialized orthant probability algorithm  of \cite{miwa2003evaluation,RSSB:RSSB625,nomura2014evaluation}.
 Unfortunately, unless a special autoregressive covariance structure is present,   these methods are hardly competitive in anything but very few dimensions. For example, the orthant algorithm of \cite{miwa2003evaluation} has complexity $\bigO(d!\times n)$, which becomes too costly for  $d>10$. For this reason, we give a comparison only with the broadly applicable   SOV scheme, which  is   widely  recognized as the current state-of-the-art method.

Since both the SOV and MET estimators are smooth, one can seek further gains in efficiency using  randomized quasi Monte Carlo. The idea behind quasi Monte Carlo is to reduce the error of the estimator by using \emph{quasirandom} or \emph{low-discrepancy} sequences of numbers, instead of the traditional \mbox{(pseudo-)} random sequences.  Typically the  error of a sample average estimator decays at the rate of $\bigO(n^{-1/2})$ when using random numbers, and at the rate of $\bigO((\log n)^d/n)$ when using pseudorandom numbers; see \cite{chopin} for an up-to-date discussion. 
 
For both the SOV and MET estimator we use the $n$-point  Richtmyer 
quasirandom sequence with randomization, as recommended by \cite{genz2009computation}.  The randomization  allows us to estimate the variability of the estimator in the standard Monte Carlo manner. The details are summarized as follows.

\begin{alg}[Randomized Quasi Monte Carlo  \citep{genz2009computation}]~
\label{Richmyer}
\begin{algorithmic}
\REQUIRE{Dimension $d$ and sample size $n$.}
\STATE{$d'\leftarrow\lceil 5d\log(d+1)/4\rceil $,\quad  $n'\leftarrow \lceil   \frac{n}{12} 
\rceil$}
\STATE{Let $p_1,\ldots,p_{d'}$ be the first $d'$ prime numbers.}
\STATE{$\bq_{i}\leftarrow \sqrt{p_i}\times(1,\ldots,n')^\top $ for  $i=1,\ldots,d'$}
\FOR{$k=1,\ldots,12$ }
\FOR{$i=1,\ldots,d-1$}  
\STATE{Let $U\sim \U(0,1)$, independently.} 
\STATE{$\mathbf{s}_i\leftarrow | 2\times[(\bq_{i} +U)\mod 1] -1 |$}
\ENDFOR
\STATE{$\mathbf{qms}\leftarrow (\bs_1,\ldots,\bs_{d-1})$}
\STATE{Use the sequence $\mathbf{qms}$ to compute an $n'$-point sample average estimator $\hat\ell_k$.}
\ENDFOR
\RETURN{ $\bar\ell\leftarrow \frac{1}{12}\sum_k\hat\ell_k$ with estimated relative error   $\frac{1}{12}\sqrt{\sum_k(\hat\ell_k-\bar\ell)^2}\Big/\bar\ell$.}
\end{algorithmic}
\end{alg}
Note that, since there is no need to integrate the $x_d$-th component, the loop over $i$ goes up to $d-1$.

 \subsection{Structured Covariance Matrices}

At this junction we assume that the matrix $A$ (or equivalently $\Sigma$) and  the bounds $\bl$ and $\bu$ have already been permuted according to the variable reordering heuristic discussed in Section~\ref{ordering}. Thus, the ordering of the variables during the integration will be the same for  both estimators and will not matter in the comparison. 

\paragraph{Example I \citep{fernandez2007perfectly}.} Consider  $\mathscr{A}=[1/2,1]^d$ with a covariance matrix
\[
\Sigma^{-1}=\frac{1}{2} I+\frac{1}{2} \mathbf{1}\mathbf{1}^\top
\] 
Columns three and four in Table~\ref{tab:1} show the estimates of $\ell$ for various values of $d$. The brackets give the estimated relative error in percentage. 
\begin{figure}[H]
\caption{Estimates of $\ell$ for various values of $d$ using $n=10^4$ replications.}
\centering
\begin{scriptsize}\setlength\extrarowheight{0pt}
\begin{tabular}{c||c|c|c|c|c|c} 
$d$ & $\ell_\mathrm{L}$ &  SOV &  MET & $\exp\left(\psi(\bx^*;\bmu^*)\right)$  & worst err. & accept pr. \\
\hline
2   & 0.0148955  &      0.0148963 ({\bf 4$\times 10^{-4}$\%}) &    0.01489 {\bf (4$\times 10^{-5}$\%)} & 0.0149     &   $2\times 10^{-4}$\%&  0.99 \\

3 & 0.0010771 &   0.0010772 ({\bf 3$\times 10^{-3}$\%}) &0.001077 {\bf(3$\times 10^{-4}$\%)} &  0.00108  &   $6\times 10^{-3}$\% &   0.99 \\

5 &  $ 2.4505\times 10^{-6}$ &   $ 2.4508\times 10^{-6}$ ({\bf 0.08\%}) &    $ 2.451\times 10^{-6}$   {  \bf(0.002\%)} &  $ 2.48\times 10^{-6}$ &   0.012\% &   0.98 \\

10 &  $ 8.5483\times 10^{-15}$ &   $   8.4591\times 10^{-15}$ ({\bf  0.8\%}) &    $ 8.556\times 10^{-15}$   {  \bf(0.01\%)} &  $  2.1046\times 10^{-14}$ &   0.03\% &    0.97 \\
15 &  $  1.3717\times 10^{-25}$ &   $   1.366\times 10^{-25}$ ({\bf  11\%}) &    $ 1.375\times 10^{-25}$   {  \bf(0.01\%)} &  $  1.43\times 10^{-25}$ &   0.04\% &    0.95 \\
20 &  $   1.7736\times 10^{-38}$ &   $    1.65\times 10^{-38}$ ({\bf  37\%}) &    $  1.7796\times 10^{-38}$   {  \bf(0.03\%)} &  $   1.869\times 10^{-38}$ &   0.05\% &     0.95\\
25 &  $    2.674\times 10^{-53}$ &   $     2.371\times 10^{-48}$ ({\bf  33\%}) &    $   2.6847\times 10^{-53}$   {  \bf(0.02\%)} &  $   2.83\times 10^{-53}$ &   0.05\% &       0.94 \\
30 &  $        6.09\times 10^{-70}$ &       - &    $   6.11\times 10^{-70}$   {  \bf(0.03\%)} &  $   6.46\times 10^{-70}$ &   0.05\% &       0.94 \\
40 &  $        2.17\times 10^{-108}$ &       - &    $   2.18\times 10^{-108}$   {  \bf(0.05\%)} &  $  2.30\times 10^{-108}$ &   0.06\% &       0.94 \\
50 &  $        2.1310\times 10^{-153}$ &       - &    $   2.1364\times 10^{-153}$   {  \bf(0.06\%)} &  $  2.24\times 10^{-153}$ &   0.05\% &       0.95\\
\end{tabular}
\end{scriptsize}
\label{tab:1}
\end{figure}
  The second column shows the  lower bound discussed in Lemma~\ref{lower} and column  five shows the deterministic upper bound.  These two bounds can then  be used to compute the exact confidence interval (mentioned in the previous section) whenever we allow $n$ to vary freely. Here, since $n$ is fixed and the error is allowed to vary, we instead display the  upper bound to the relative error 
		(given in column six under the ``worst err.'' heading)
	\[\sqrt{\Var(\bar\ell)}/\ell\leq (\exp\left(\psi(\bx^*;\bmu^*)\right)/\ell_\mathrm{L}-1)/\sqrt{n}.
	\]
 Finally, column seven (accept pr.) gives the acceptance rate of  Algorithm~\ref{accept-reject}
when using the instrumental density $g(\cdot\,;\bmu^*)$ with enveloping constant
$c=\exp\left(\psi(\bx^*;\bmu^*)\right)$. 

What makes the MET approach better than other methods?
First, the acceptance rate in column seven remains high even for $d=50$. In contrast, the acceptance rate from  naive acceptance-rejection with
instrumental pdf $\phi(\mathbf{0},\Sigma)$ is a rare-event probability of approximately $2.13\times 10^{-153}$. Note again that the existing accept-reject scheme  of \cite{chopin2011fast} is an excellent algorithm designed for extremely fast simulation in one or two dimensions (in quite general settings) and is not suitable here. 


 Second, the performance of both the SOV and MET estimators   
gradually deteriorates with increasing $d$. However, the SOV 
estimator  has  larger relative error, does not give meaningful 
results for $d>25$, and possesses no theoretical quantification of 
its performance. In contrast, the MET estimator is  guaranteed to 
have better relative error than the one  given in column six (worst 
err.).

Finally, in further numerical experiments (not displayed here) we 
 observed that the width, $\epsilon$, of the \emph{exact} confidence 
interval, 
$\bar\ell\pm\epsilon$ with  $\alpha=0.05$, based on the Hoeffding 
bound \eqref{Hoeffding}, 
was of the same order of magnitude as the width of the \emph{
approximate} confidence interval $\bar\ell\pm z_{1-\alpha/2}\times S/
\sqrt{n(\epsilon)}$.

\paragraph{Example II  \citep{fernandez2007perfectly}.}
Consider the hypercube $\mathscr{A}=[0,1]^d$ and the isotopic covariance with elements
\[
(\Sigma^{-1})_{i,j}=\frac{1}{2^{|i-j|}}\times\I\{|i-j|\leq d/2 \}\;.
\]
\vspace{-0.9cm}
\begin{figure}[H]
\caption{Estimates of $\ell$ for various values of $d$ using $n=10^4$ replications.}
\centering
\begin{scriptsize} \setlength\extrarowheight{0pt}
\begin{tabular}{c||c|c|c|c|c|c}
$d$ & $\ell_\mathrm{L}$ &  SOV & MET & $\exp\left(\psi(\bx^*;\bmu^*)\right)$  & worst err. & accept pr. \\
\hline

2   & 0.09114  &        0.09121 ({\bf 6$\times 10^{-4}$\%}) &     0.09121 {\bf (2$\times 10^{-4}$\%)} &  0.09205   & $  0.009\%$&  0.99 \\

3   & 0.02303  &        0.02307 ({\bf 0.001\%}) &    0.02307 {\bf (4$\times 10^{-4}$\%)} &   0.0234     & $  0.01\%$&  0.98 \\

10 &  $     1.338\times 10^{-6}$ &     $1.3493\times 10^{-6}$   {  \bf(0.03\%)}  &    $        1.3490\times 10^{-6}$   { (\bf 0.003\%)} &  $    1.454\times 10^{-6}$ &      0.07\% &         0.92\\

20 &  $    1.080\times 10^{-12}$ &    $1.0982\times 10^{-12}$   {  \bf(0.23\%)}  &    $       1.0989\times 10^{-12}$   { (\bf 0.004\%)} &  $   1.289\times 10^{-12}$ &     0.17\% &         0.85 \\
25 &  $   9.770\times 10^{-16}$ &     $1.00\times 10^{-15}$   {  \bf(0.28\%)}  &    $       9.9808\times 10^{-16}$   { (\bf 0.02\%)} &  $   1.222\times 10^{-15}$ &     0.2\% &         0.81 \\
50 &  $    5.925\times 10^{-31}$ &     $6.137\times 10^{-31}$   {  \bf(0.7\%)}  &    $       6.188\times 10^{-31}$   { (\bf 0.05\%)} &  $  9.368\times 10^{-31}$ &     0.5\% &         0.66 \\
80 &  $    3.252\times 10^{-49}$ &     $3.477\times 10^{-49}$   {  \bf(1.8\%)}  &    $      3.479\times 10^{-49}$   { (\bf 0.1\%)} &  $   6.812\times 10^{-49}$ &     1.0\% &         0.50 \\
100 &  $    2.18\times 10^{-61}$ &     $2.351\times 10^{-61}$   {  \bf(3\%)}  &    $      2.384\times 10^{-61}$   { (\bf 0.2\%)} &  $   5.50\times 10^{-61}$ &     1.3\% &         0.43 \\
120 &  $    1.462\times 10^{-73}$ &     $1.641\times 10^{-73}$   {  \bf(5.6\%)}  &    $      1.622\times 10^{-73}$   { (\bf 0.3\%)} &  $   4.45\times 10^{-73}$ &     1.7\% &         0.36 \\
150 &  $    8.026\times 10^{-92}$ &     $9.751\times 10^{-92}$   {  \bf(6.3\%)}  &    $     9.142\times 10^{-92}$   { (\bf 0.18\%)} &  $   3.23\times 10^{-91}$ &     2.5\% &         0.28 \\
200 &  $    2.954\times 10^{-122}$ &     $3.581\times 10^{-122}$   {  \bf(11\%)}  &    $    3.525\times 10^{-122}$   { (\bf 0.5\%)} &  $   1.905\times 10^{-121}$ &     4.4\% &         0.18 \\
250 &  $1.087\times 10^{-152}$ &     $ 1.359\times 10^{-152}$   {  \bf(15\%)}  &    $    1.357\times 10^{-152}$   { (\bf 0.6\%)} &  $    1.120\times 10^{-151}$ &     7.2\% &         0.12 \\
\end{tabular}
\end{scriptsize}
\end{figure}
Observe how rapidly the probabilities become very small.
Why should we be interested in estimating small ``rare-event'' probabilities? The simple answer is that
all probabilities  become eventually rare-event probabilities as the dimensions get larger and larger, making naive accept-reject simulation infeasible.  These small probabilities sometimes present not only theoretical challenges (rare-event estimation), but  practical ones like representation in finite precision arithmetic and numerical underflow. For instance, in using the SOV estimator \cite{RSSC:RSSC1007}  note that:
\emph{``Numerical
problems arise for very small probabilities, e.g. for observations from different components. To avoid these problems observations with a small posterior probability (smaller than or equal to $10^{-6}$) are omitted in the M-step of this component."}
The  MET estimator is not immune to numerical underflow and loss of precision during computation, but consistent with Theorems~\ref{BRE} and \ref{VRE}, it is typically much more robust than the SOV estimator in estimating small  probabilities.

\subsection{Random Correlation Matrices}
One can argue that the covariance matrices we have considered so far are too structured and hence not representative of a ``typical" covariance matrix. Thus,
for simulation and testing \cite{miwa2003evaluation}  and \cite{RSSB:RSSB625} find it desirable to  use random correlation matrices. In the subsequent examples we  use   the method of \cite{davies2000numerically} to simulate random test correlation matrices whose  eigenvalues  are  uniformly distributed over the simplex $\{\bx: x_1+\cdots+x_d=d\}$.

%

\paragraph{Example III.} A natural question is whether the MET estimator would  still be preferable
when integrating over a ``non-tail'' region such as $\mathscr{A}=[-1/2,\infty]^{100}$.   The table below summarizes the output of running the algorithms on 100 independently simulated random correlation matrices. Both the SOV and MET estimators  used $n=10^5$ quasi Monte Carlo points. The `accept rate' row displays the five number summary of the  estimated acceptance probability of Algorithm~\ref{accept-reject}. 

\begin{figure}[H]
\caption{Table: five number summary for relative error   based on 100 independent replications; Graph: boxplots of these 100 outcomes on  logarithmic scale. }
\vspace{0.5cm}
\centering \setlength\extrarowheight{2pt}
{\small\begin{tabular}{c||c|c|c|c|c}
  & min  &       1-st quartile &   median &  3-rd quartile   &  max\\
\hline
 MET&   0.07\%   &  0.12\%   &  0.17\% &    0.20\% &    0.44\%\\
  SOV&    0.27\% &    0.63\%   &   1.00\%   &    1.68\%  &   9.14\%\\
	\hline
    accept rate&1.2\%   &  3.9\%  &   5.5\%   &   7.3\%   &    12\%
\end{tabular}
}
\includegraphics[scale=0.7]{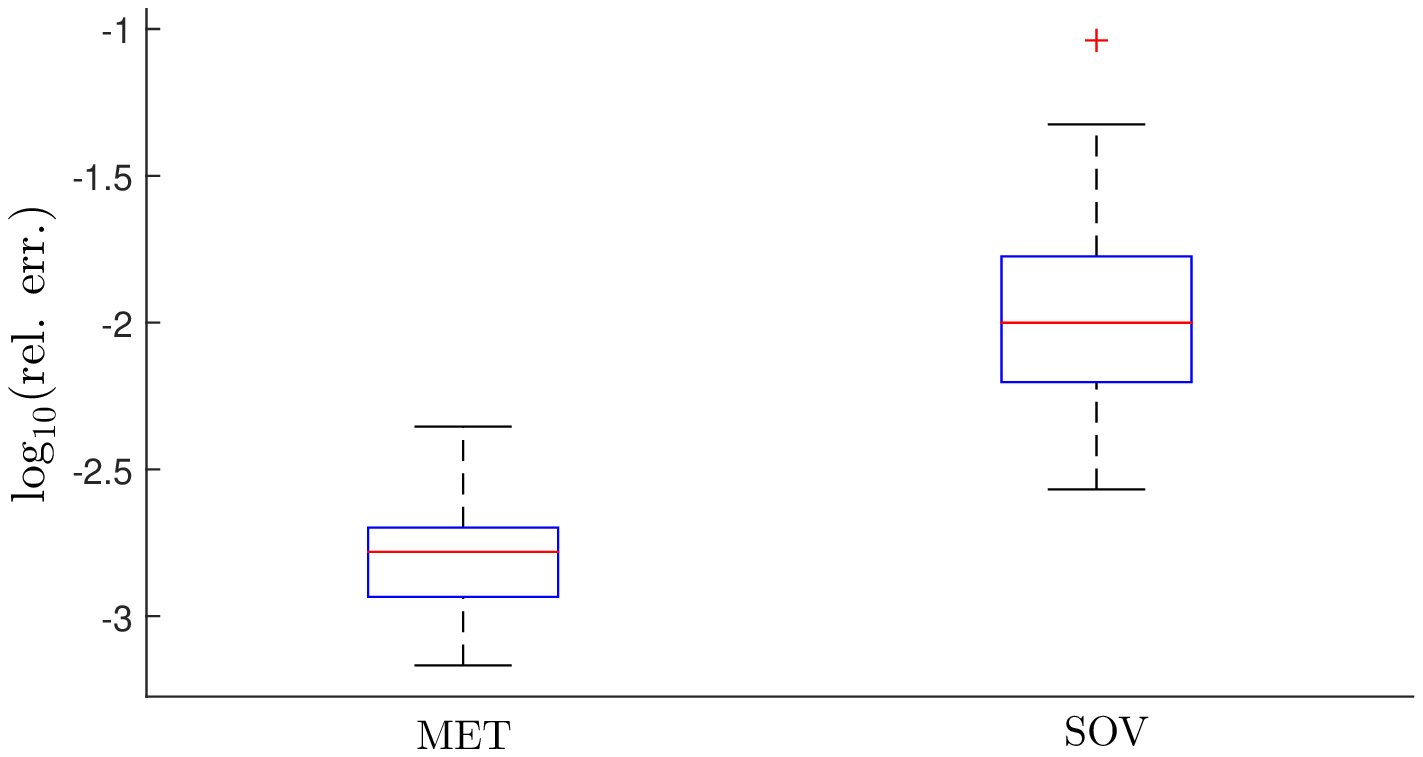}
\label{fig:orthant}
\end{figure}

%

So far we have said little about the cost of computing the optimal pair $(\bx^*;\bmu^*)$, and
the measures of efficiency we have considered  do not account for the computational cost of the estimators.  The reason for this is that in  the examples we investigated, the computing time required to find the pair $(\bx^*;\bmu^*)$ 
 is insignificant  compared to the time it takes to evaluate $n>10^5$ replications of $\hat\ell$ or $\mathring{\ell}$. 


In the current example, the numerical experiments suggest that the MET estimator is roughly 20\% more costly than the SOV estimator. If one adjusts the results in Figure~\ref{fig:orthant} in order to account for this  time difference, then the relative error in the SOV row would  be reduced by a factor of at most $1.2$. This adjustment will thus give a reduction in the typical (median) relative error from $1.0$ to $1/1.2\approx 0.83$ percent, which is hardly significant.


%
%
%
%
\paragraph{Example IV.} Finally, we wish to know if the strong efficiency  described in Theorem~\ref{BRE} 
may benefit the MET estimator as we move further into the tails of the distribution. 
Choose the ``tail-like'' $\mathscr{A}=[1,\infty]^{100}$  and use $n=10^5$. The following table and graph summarize the results of 100 replications.
\begin{figure}[H]
\caption{Relative errors of SOV and MET estimators over 100 random correlation cases.}
\vspace{0.5cm}
\centering  \setlength\extrarowheight{2pt}
{\small
\begin{tabular}{c||c|c|c|c|c}
 & min  &       1-st quartile &   median &  3-rd quartile   &  max\\
\hline
MET&  0.020\%  &   0.044\%  &   0.077\%   &   0.12\%  &    0.44\%\\
     SOV&  4.3\%   &    15\%  &     26\%   &    48\%  &     99\%\\
		\hline
     accept rate  &  1.5\% &  10\%    &    18\%    &     26\%   &  43\%
\end{tabular}
}
\includegraphics[scale=0.7]{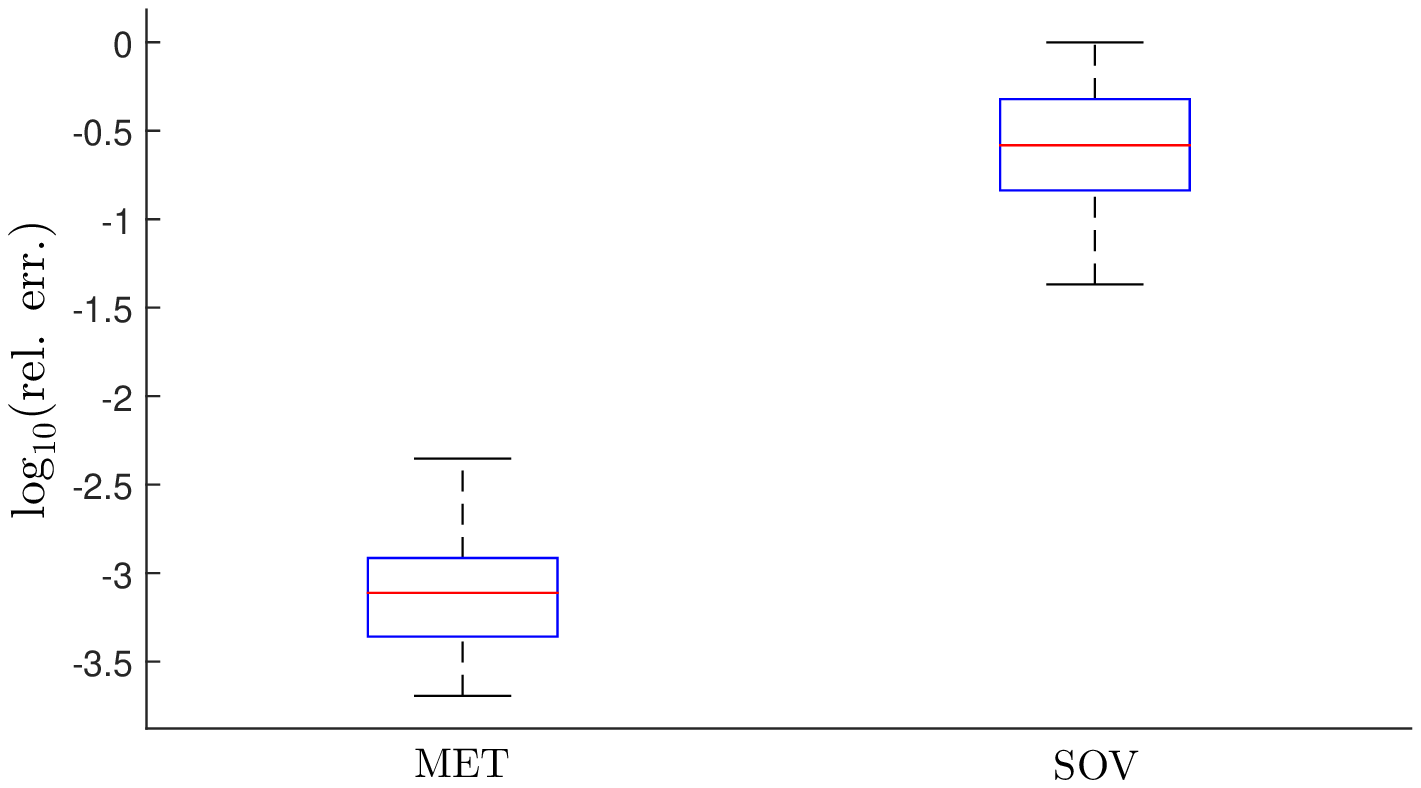}
\end{figure}
As seen from the results, in this particular example the variance  of the MET estimator is typically more than $10^5$ times smaller than the variance of the SOV estimator.


\subsection{Computational Limitations In High Dimensions}
\label{reordering}
It is important to emphasize the limitations of the minimax  tilting approach. Like all other methods, including MCMC, it is not a panacea against the curse of dimensionality. The acceptance probability of Algorithm~\ref{accept-reject} ultimately becomes a rare-event probability as the dimensions keep increasing, because the bounded or vanishing relative error properties of $\hat\ell$ do not hold in the asymptotic regime $d\uparrow\infty$. 

 Numerical experiments suggest that the method generally works reliably for $d\leq 100$. The approach may sometimes be effective in higher dimensions provided $\ell$
does not decay too fast in $d$. In this regard, \cite{miwa2003evaluation,RSSB:RSSB625} study the orthant probability $\ell=\Pm(\bX\in [0,\infty]^d)$ with the positive correlation structure
\[
\Sigma=\frac{1}{2} I+\frac{1}{2}\mathbf{1}\mathbf{1}^\top\;.
\]
This is a rare case for which the exact value of the probability is known, namely $\ell= 1/(d+1)$, and decays very slowly to zero as $d\uparrow\infty$. For this reason, we use it 
to illustrate the behavior of the SOV and MET estimators for very large $d$.

\begin{figure}[H]
\caption{Graph: computational cost  in seconds; Table: relative error  in percentage;}
\label{fig:cost}
\begin{minipage}{0.65\linewidth}
 \hspace{-1cm} \includegraphics[scale=0.5]{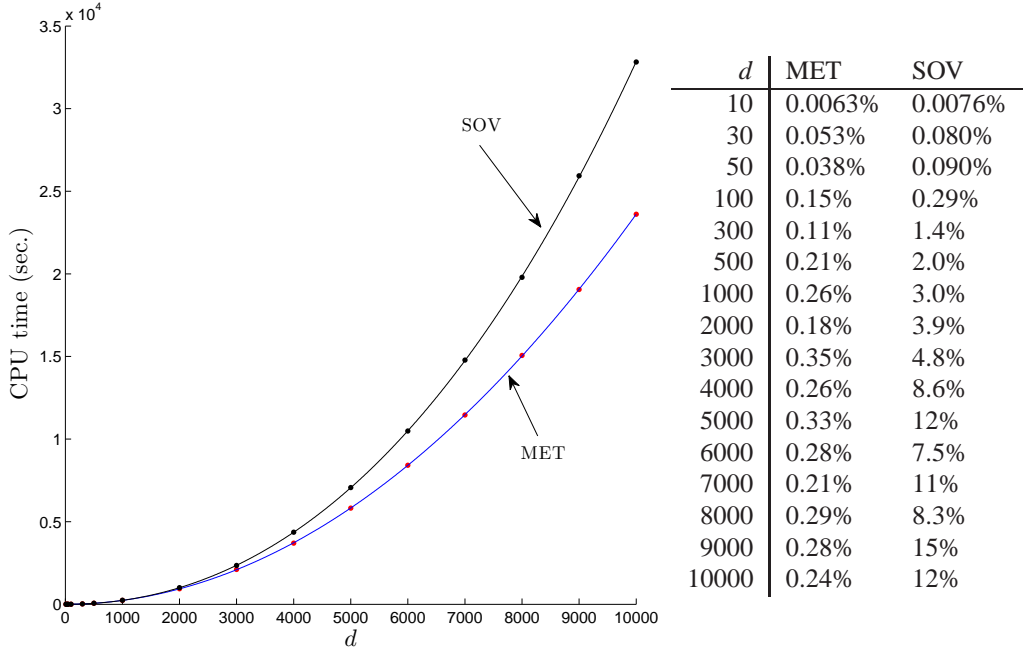}
 \end{minipage}~\begin{minipage}{0.3\linewidth}
{\small
\begin{tabular}{r|ll}
$d$& MET & SOV \\
\hline
           10  &  0.0063\%   & 0.0076\%\\
           30  &   0.053\%   &  0.080\%\\
           50   &  0.038\%   &  0.090\%\\
          100   &   0.15\%   &   0.29\%\\
          300   &   0.11\%   &    1.4\%\\
          500   &   0.21\%   &    2.0\%\\
         1000   &   0.26\%   &    3.0\%\\
         2000   &   0.18\%   &    3.9\%\\
         3000   &   0.35\%   &    4.8\%\\
         4000   &    0.26\%   &    8.6\%\\
         5000   &   0.33\%   &     12\%\\
         6000  &   0.28\%    &   7.5\%\\
         7000  &     0.21\%   &    11\%\\
         8000  &    0.29\%   &     8.3\%\\
         9000  &    0.28\%   &    15\%\\
        10000  &    0.24\%   &    12\%
\end{tabular}
}
 \end{minipage}
\end{figure}

  Figure~\ref{fig:cost} shows the output of a numerical experiment with $n=10^5$
for various values of $d$. The graph on the left gives the computational cost
in seconds. Both the SOV and the MET estimators have  cost 
of $\bigO(d^3)$ --- hence the excellent agreement with the
 least squares cubic polynomials fitted to the empirical CPU 
data. The table on the right displays the  relative error  
for both methods. In this example, we apply the variable 
reordering heuristic to the SOV estimator only, illustrating 
that  the heuristic is not always necessary to achieve satisfactory performance with the MET estimator.

This example confirms the result  in Theorem~\ref{VRE} that the SOV estimator works better in settings with strongly positive correlation structure (but poorly with negative correlation). Further,  the results suggest the MET estimator is also aided by the presence of positive correlation.

\subsection{Exact Simulation of Probit Posterior}
A popular GLM \citep{koop2007bayesian}  for  binary responses $\by=(y_1,\ldots,y_m)^\top$ with
explanatory variables $\bx_i=(1,x_{i2},\ldots,x_{ik})^\top,\;i=1,\ldots,m$ is the probit Bayesian model:
\begin{itemize}
\item Prior: $p(\bb)\varpropto \exp\left( -\frac{1}{2}(\bb-\bb_0)^\top V^{-1}  (\bb-\bb_0) \right)$ with $\bb\in\mathbb{R}^k$ 
and   for simplicity  $\bb_0=\mathbf{0}$;
\item Likelihood: $p(\by\gvn \bb)\varpropto \exp\left(\sum_{i=1}^m\log\Phi\Big((2y_i-1)\bx_i^\top\bb\Big)\right)$.
\end{itemize}
 The challenge is to simulate from the posterior $p(\bb\gvn\by)$.
One can use  latent variables \citep{albert1993bayesian}  to  represent the 
posterior as the marginal of a truncated multivariate normal. 
Let $\blam\sim\Nor(\mathrm{0},I_m)$ be latent variables and define the design matrix $\tilde X=\mathrm{diag}(2\by-\mathbf{1})X$. 
Then,  the marginal $f(\bb)$ of the joint pdf 
\[
\textstyle
f(\bb,\blam)\varpropto \exp\left( -\frac{1}{2}\|V^{-1/2}\bb\|^2-\frac{1}{2}\|\blam\|^2 \right) 
\I\{\tilde X\bb-\blam\geq \mathbf{0} \}
\]
equals the desired posterior $p(\bb\gvn \by)$. We can thus apply our accept-reject scheme, because the joint $f(\bb,\blam)$ is  of the desired truncated multivariate  form \eqref{prob} with  $d=k+m$ and
\[\textstyle
\bz=\left[\begin{array}{c}
V^{-1/2}\bb\\
\blam\end{array}\right],\quad A=\left(\tilde XV^{1/2}, -I\right),\quad \bl=\mathbf{0},\quad \bu=+\boldsymbol\infty.
\]

\begin{figure}[H]
\caption{Marginal distribution of $\bb$ computed from 8000 exact iid realizations.}
\centering{
\includegraphics[scale=.81]{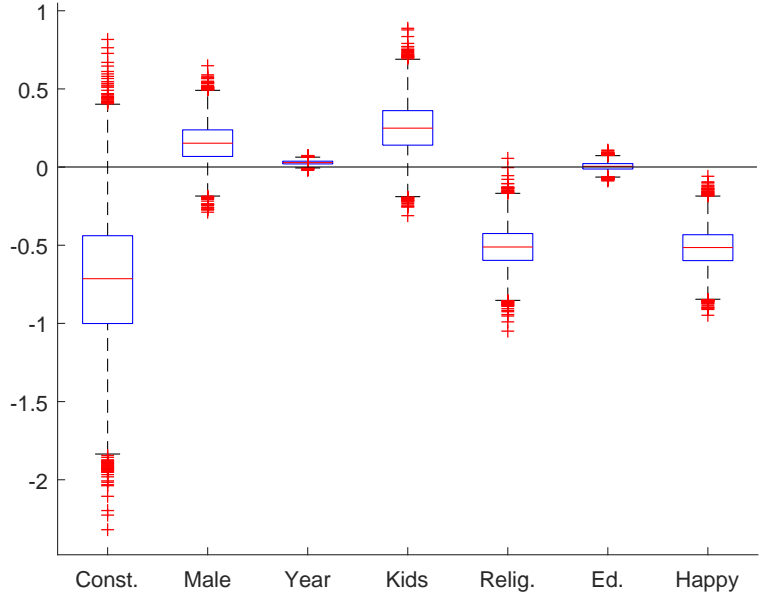}
}
\label{affairs}
\end{figure}

As an numerical example, we apply the probit model to the widely studied
\emph{extramarital affairs} dataset from \cite{koop2007bayesian}. The dataset contains $m=601$ independent observations: the binary response $y_i$ indicates if the 
$i$-th respondent has had an extramarital affair; the six explanatory variables ($k=7$) are
male indicator (Male), number of years married (Year), `has' or `has not' children (Kids), religious or not (Relig.), years of formal education (Ed.), and a binary variable denoting whether the marriage is happy or not (Happy). Figure~\ref{affairs} shows the boxplots of the marginal distributions of $\beta_1,\ldots,\beta_7$ based on $8000$ iid simulations from the posterior $p(\bb\gvn\by)$ with prior covariance $V=5 I$.

The conclusion that only years of marriage, religiosity, and conjugal happiness are statistically significant is, of course, well known \citep{koop2007bayesian} and used to validate our new simulation scheme. The question is what have we gained in using minimax tilting?

On the one hand, for the first time we have conducted the Bayesian inference using exact iid samples from the  posterior and we did not have to fret about unquantifiable  issues such as `burn-in' and `mixing-speed' as is typical with approximate MCMC simulation \citep{philippe2003perfect}. 

On the other hand, the acceptance rate in the simulation was  $1/217$, that is, we had to simulate (on average) 217 random vectors to accept one as an exact independent realization from the posterior. Admittedly, this acceptance rate  could have been better and as shown in the previous experiments it is going to deteriorate with increasing dimensionality. However, there are hardly any alternatives for exact sampling --- naive acceptance rejection  for the extramarital data would enjoy an acceptance rate of $\bigO(10^{-146})$ and without minimax tilting (say, with proposal  $g(\bx;\mathbf{0})$)
the Accept-Reject Algorithm~\ref{accept-reject}  enjoys an acceptance rate of  $\bigO(10^{-16})$.

Thus,  our main point stands: the proposed accept-reject scheme 
can be used for exact simulation whenever, say $d\leq 100$, and when $d$ is in the thousands it can be used to 
accelerate  Gibbs sampling by  grouping or blocking dozens of highly correlated variables together  \citep{chopin2011fast,philippe2003perfect}.

\section*{Concluding Remarks}

The minimax tilting method can be effective for exact simulation from
the truncated multivariate normal distribution. The proposed method permits us to  dispense with Gibbs sampling in   dimensions less than 100, and for larger dimensions to accelerate 
existing Gibbs samplers by sampling jointly hundreds of highly correlated variables. 

The minimax approach can also be used to estimate normal probability integrals. Theoretically, the method improves on the already excellent SOV estimator and in a tail asymptotic regime it can achieve the best possible efficiency --- vanishing relative error. The numerical experiments suggest that the proposed method can be significantly more accurate than the widely used SOV estimator, especially in the tails of the distribution. The experiments also point out to its  limitations --- as the dimensions get larger and larger  it eventually  fails. 

The minimax tilting approach  in this article  can be extended to 
other multivariate densities related to the normal.
 Upcoming work by the author will argue that  significant efficiency gains are also possible  in the case of the multivariate student-$t$   and  general elliptic distributions for which a strong log-concavity property holds. Just as in the multivariate normal case, the approach permits us  to estimate accurately hitherto intractable student-$t$ probabilities, for which existing estimation schemes exhibit relative error close to 100\%.


\section*{Acknowledgments}
This work was supported by the  Australian Research Council   under grant DE140100993.

\appendix
\section{Appendix}
\subsection{Proof of Lemma~\ref{lemma}}

First, we show that   $\psi$ is a concave function of $\bx$ for any $\bmu$. To see this, note that
if $Z\sim \Nor(0,1)$ under $\Pm$, then  by the well-known properties of log-concave measures \citep{prekopa}, the function $q_1:\mathbb{R}\rightarrow \mathbb{R}$ defined as
\[\textstyle
q_1(w)=\log\Pm( l\leq Z+w\leq u)=\log\frac{1}{\sqrt{2\pi}}\int_\mathbb{R} \exp\left(-\frac{1}{2}z^2\right) \I_{\{ (Z+w)\in \mathscr{Z}\}} \di z\;,
\] where $\mathscr{Z}=[l,u]$ is a convex set,
is a concave function of $w\in \mathbb{R}$. Hence,  for an arbitrary linear map $C\in\mathbb{R}^{d\times 1}$, the function
$q_2: \mathbb{R}^{d}\rightarrow\mathbb{R}$ defined as $q_2(\bx)=q_1(C\bx)$  is concave as well. 
It follows that each function 
\[
\log\Pm(\tilde l_k\leq Z+\mu_k\leq \tilde u_k)=\log\Pm((Z+C_k\bx)\in \mathscr{Z}_k)
\]
 (using the obvious choices of $C_k$ and $\mathscr{Z}_k$) is concave in $\bx$.
Hence,  $\psi$ is concave in $\bx$, because it is a non-negative weighted sum of concave functions.

Second, we show that $\psi$ is convex in $\bmu$ for each value of $\bx$. After some simplification, we can write
\[
\psi(\bx;\bmu)=-\bx^\top\bmu+\sum_k \log\Em\,\exp\left(\mu_k Z\right)\I_{\{\tilde l_k\leq Z\leq \tilde u_k\}}\;.
\] Now, each
of $\log\Em\,\exp(\mu_k Z)\I_{\{\tilde l_k\leq Z\leq \tilde u_k\}}$ is  convex in $\mu_k$, because up to a normalizing constant, this is the cumulant generating function of a standard normal random variable $Z$, truncated to $[\tilde l_k,\tilde u_k]$. Since a non-negatively weighted sum of convex functions is convex, we conclude that $\psi(\bx;\bmu)$ is convex in $\bmu$. Finally, since convexity is preserved under pointwise supremum, $\sup_{\bx\in \mathscr{C}}\psi(\bx;\bmu)$ is still convex in $\bmu$. Moreover,  here we have the strong min-max property:
$
\inf_{\bmu}\sup_{\bx\in \mathscr{C}}\psi(\bx;\bmu)=\sup_{\bx\in \mathscr{C}}\inf_{\bmu}\psi(\bx;\bmu),
$ from which the lemma follows.
$\hfill\Box$

\subsection{Proof of Theorem~\ref{BRE}}
\label{proof:theorem}

Before proceeding with the proof we  note the following.

First, using the necessary and sufficient condition \eqref{kkt}, we can write the solution of \eqref{qqp} explicitly
as $\bx_1=\gamma L_{11}^{-1}\pl_1,\;\bx_2=\mathbf{0}$ with minimum
$\frac{\gamma^2}{2}\|L_{11}^{-1}\pl_1\|^2$. In addition, from \eqref{kkt}
we can also deduce that $\blam_1=\gamma L_{11}^{-\top}L_{11}^{-1}\pl_1>\mathbf{0}$ and 
$
\bq= L_{21}L_{11}^{-1}\pl_1-\pl_2\geq\mathbf{0}.
$

Second,  the asymptotic behavior of $\ell(\gamma)=\Pm(\bX\geq\gamma\bl)$ has been established
by  \cite{hashorva2003multivariate}. For convenience, we restate  their  result  using  our simplified notation.

\begin{proposition}[\cite{hashorva2003multivariate}]
\label{mills refined}
Consider the tail probability $\ell(\gamma)=\Pm(\bX\geq \gamma\bl)$, where $\bX\sim\Nor(\mathbf{0},\Sigma)$ and $\gamma>0,\; \bl>\mathbf{0}$.
Define the set $\sJ$ as in \eqref{J}.
Then, the  tail behavior of $\ell(\gamma)$ as $\gamma\uparrow\infty$ is 
\[
\ell(\gamma)\simeq c\times \exp\left(-\frac{\gamma^2}{2}\|L_{11}^{-1}\pl_1\|^2 -\sum_{k=1}^{d_1}\log \left(\gamma \Big\{L_{11}^{-\top}L_{11}^{-1}\pl_1\Big\}_k \right)\right),
\]
where the constant $c$ is given by:
\[
c=\frac{\Pm(Y_{j}>0,\forall j\in \sJ)}{(2\pi)^{d_1/2}|L_{11}|},\quad (Y_1,\ldots,Y_{d_2})^\top\sim\Nor(\mathbf{0},L_{22}L_{22}^\top)
\]
if $\sJ\not=\emptyset$, and $c=(2\pi)^{-d_1/2}|L_{11}|^{-1}$ if $\sJ=\emptyset$.
\end{proposition}

The last two observations pave the way to proving that, depending on the set $\sJ$, either
$\exp(\psi(\bx^*,\bmu^*))=\bigO(\ell(\gamma))$, or
$\exp(\psi(\bx^*,\bmu^*))\simeq\ell(\gamma)$. The details of the argument are as follows.

In the setting of Theorem~\ref{BRE}, the  Karusch-Kuhn-Tucker conditions \eqref{KKT}  simplify to:
\begin{equation}
\label{kkt2}
\begin{split}
\bmu-\bx+\bPsi&=\mathbf{0}\\
-\bmu+(\breve L^\top-I)\bPsi+\breve L ^\top\bfeta&=\mathbf{0}\\
\bfeta\geq \mathbf{0},\;\;\gamma\pl-L\bx&\leq\mathbf{0}\\
\bfeta^\top(\gamma\pl-L\bx)&=0
\end{split}
\end{equation}
where $\bfeta$ is a  Lagrange multiplier (corresponding to $\bfeta_2$ in
\eqref{KKT}) and we replaced $\bl$ with  $\gamma\pl$. 
\paragraph{Case $\sJ=\emptyset$.}
We  now verify by substitution that, if $\sJ=\emptyset$, the unique solution of  \eqref{kkt2} is of the  asymptotic form
\begin{equation}
\label{asym}
\begin{split}
\bx_1&\simeq\tilde \bx_1=\gamma L_{11}^{-1}\pl_1\\
 \bx_{2}&\simeq \tilde \bx_{2}=o(\mathbf{1})\\
 \bmu_1&\simeq \tilde\bmu_1=- \gamma(D_1L_{11}^{-\top}-I)L_{11}^{-1}\pl_1\\
 \bmu_{2}&\simeq \tilde\bmu_{2}=o(\mathbf{1}) \\
 \bfeta&\simeq\tilde\bfeta= o(\mathbf{1})
\end{split}
\end{equation}
Equation four in \eqref{kkt2} is obviously satisfied, because $\tilde\bfeta$ tends to zero by assumption in \eqref{asym}. Next, note that $-\gamma\left(L_{21}L_{11}^{-1}\pl_1-\pl_2\right)-L_{22}\tilde\bx_2=-\gamma\bq
+o(\mathbf{1})\downarrow-\boldsymbol\infty$, as $\gamma\uparrow\infty$. Hence, line three in \eqref{kkt2} is also satisfied for sufficiently large $\gamma$:
\[
\gamma\pl-L\tilde\bx=\left(\begin{array}{cc}
\gamma\pl_1- L_{11} \tilde\bx_1\\
\gamma\pl_2-L_{21}\tilde\bx_1-L_{22}\tilde\bx_2
\end{array}\right)=
\left(\begin{array}{cc}
\mathbf{0}\\
-\gamma\left(L_{21}L_{11}^{-1}\pl_1-\pl_2\right)
-L_{22}\tilde\bx_2\end{array}\right).
\]
Next, note that
\[
\begin{split}
\tilde \bl_1&=D^{-1}_1(\gamma\pl_1-( L_{11}-D_1)\tilde\bx_1)=\gamma  L_{11}^{-1}\pl_1=\tilde\bx_1\\
\tilde \bl_2&=D^{-1}_2(\gamma\pl_2- L_{21}\tilde\bx_1-(L_{22}-D_2)\tilde\bx_2)=
-\gamma D^{-1}_2\bq+o(\mathbf{1})\downarrow-\boldsymbol\infty
\end{split}
\]
Hence, from 
$
\tilde\bl_1-\tilde\bmu_1= \gamma  L_{11}^{-1}\pl_1+\gamma(D_1L_{11}^{-\top}-I)L_{11}^{-1}\pl_1=\gamma D_1L_{11}^{-\top}L_{11}^{-1}\pl_1
=D_1\blam_1>\mathbf{0}$ and
$
\tilde\bl_2-\tilde\bmu_2=-\gamma D_{2}^{-1}\bq +o(\mathbf{1}),
$ and Mill's ratio ($\phi(\gamma;0,1)/\overline\Phi(\gamma)\simeq\gamma$  and $\phi(-\gamma;0,1)/\overline\Phi(-\gamma)\downarrow 0$)
 we obtain the asymptotic behavior of $\bPsi$:
\[
\begin{split}
\bPsi_1&\simeq  \gamma D_1 L_{11}^{-\top}L_{11}^{-1}\pl_1,\qquad
\bPsi_{2}=o(\mathbf{1})\;,
\end{split}
\]
where we recall that $\blam_1=\gamma L_{11}^{-\top}L_{11}^{-1}\pl_1>\mathbf{0}$.
Equation one  in \eqref{kkt2} thus simply verifies that
\[
\begin{split}
\tilde\bx_1&=\bPsi_1+\tilde\bmu_1\simeq   \gamma D_1 L_{11}^{-\top}L_{11}^{-1}\pl_1- \gamma(D_1L_{11}^{-\top}-I)L_{11}^{-1}\pl_1=\gamma L_{11}^{-1}\pl_1\\
\tilde\bx_{2}&=\bPsi_{2}+\tilde\bmu_{2}=o(\mathbf{1})
\end{split}
\]
Equation one and two yield $\bx=\breve L^\top\bPsi =L^\top D^{-1}\bPsi$, which again is easily verified:
\[
\begin{split}
\bx_1&= L_{11}^\top D^{-1}_1\bPsi_1+L_{21}^{\top} D^{-1}_2\bPsi_2\simeq  \gamma L_{11}^{-1}\pl_1=\tilde\bx_1\\
\bx_2&=  L_{22}^\top D_2^{-1}\bPsi_2=o(\mathbf{1})=\tilde \bx_2
\end{split}
\]
The asymptotic behavior of  $\psi^*=\psi(\bx^*;\bmu^*)$ is obtained 
by evaluating $\psi$ at the  asymptotic solution \eqref{asym}, that is,
$\tilde\psi\idef\psi(\tilde\bx;\tilde\bmu)=$
\begin{equation}
\label{manipulation}
\begin{split}
&=\frac{\|\tilde\bmu\|^2}{2}-\tilde\bx^\top\tilde\bmu+\sum_{k=1}^d\log\overline\Phi(\tilde l_k-\tilde\mu_k),\qquad \textrm{where by definition }\; \overline\Phi(x)\idef\Pm(Z>x) \\
&= \frac{\|\tilde\bmu_1\|^2}{2}-\tilde\bx_1^\top\tilde\bmu_1+\bigO(\|\tilde\bx_2\|^2)+\sum_{k=1}^{d_1}\log\overline\Phi(\tilde l_k-\tilde\mu_k)+\sum_{k=1}^{d_2} \log\overline\Phi(-\gamma \{D_{2}^{-1}\bq\}_{k}+o(1)))
\end{split}
\end{equation}
It follows from  Mill's  ratio, $\log\overline\Phi(\gamma)\simeq -\frac{1}{2}\gamma^2-\log\gamma-\frac{1}{2}\log(2\pi)$,  and $\log\overline\Phi(-\gamma)\uparrow 0$ that
\[
\begin{split}
\tilde\psi&= \frac{\|\tilde\bmu_1\|^2}{2}-\tilde\bx_1^\top\tilde\bmu_1-
\frac{\gamma^2}{2} \|D_1L_{11}^{-1} L_{11}^{-1}\pl_1\|^2-\frac{d_1}{2}\log(2\pi)-\sum_{k=1}^{d_1}\log(\gamma \{D_1L_{11}^{-1} L_{11}^{-1}\pl_1\}_k)+o(1) \\
& = -\frac{\gamma^2}{2}\|L_{11}^{-1}\pl_1\|^2-\frac{d_1}{2}\log(2\pi)-\log|L_{11}| -\sum_{k=1}^{d_1}\log(\gamma \{L_{11}^{-1} L_{11}^{-1}\pl_1\}_k)+o(1)
\end{split}
\]
In other words, from
 Proposition~\ref{mills refined} we have that $\exp(\tilde\psi)\simeq\ell(\gamma)$ as $\gamma\uparrow\infty$.  Therefore,  
\begin{equation*}
\begin{split}
\frac{\Var_{\bmu^*}(\hat\ell)}{\ell^2}= \frac{\Em_{\bmu^*} \exp(2\psi(\bX;\bmu^*))}{\ell^2}-1&\leq \frac{\exp(\psi(\bx^*;\bmu^*))\Em_{\bmu^*} \exp(\psi(\bX;\bmu^*))}{\ell^2}-1\\
&\leq \frac{\exp(\psi(\bx^*;\bmu^*))}{\ell(\gamma)}-1 \simeq \frac{\exp(\tilde\psi)}{\ell(\gamma)}-1=o(1)\;.
\end{split}
\end{equation*}

It follows that for $\sJ=\emptyset$ the minimax  estimator \eqref{tilted est} exhibits vanishing relative error --- the best possible  asymptotic tail behavior.

\paragraph{Case $\sJ\not=\emptyset$.}
 Recall that $(\breve\bx,\breve\bmu)$ is the solution of the 
nonlinear system \eqref{nonlin}, as well as the optimization 
program \eqref{optim} without its constraint $\bx\in\mathscr{
C}$ (note that a reordering of the variables via the 
permutation matrix $P$ does not change the statement of 
\eqref{optim}  or \eqref{nonlin}). We have $\psi(\bx^*;\bmu^*
)\leq \psi(\breve\bx;\breve\bmu)$, because dropping a 
constraint in the maximization of \eqref{optim} cannot 
reduce the maximum.
As in the case of $\sJ=\emptyset$, one can then  verify via direct substitution that 
\[
\tilde\bx_1=\gamma L_{11}^{-1}\pl_1,\quad \tilde\bx_2=\bigO(\mathbf{1}),\quad
\tilde\bmu_1=- \gamma(D_1L_{11}^{-\top}-I)L_{11}^{-1}\pl_1,\quad \tilde\bmu_2=\bigO(\mathbf{1})
\]
is the asymptotic form of the solution to \eqref{nonlin}. In other words,
$\tilde\psi=\psi(\tilde\bx;\tilde\bmu)\simeq\psi(\breve\bx;\breve\bmu)\geq \psi(\bx^*;\bmu^*)$. Similar manipulations as the ones in \eqref{manipulation}  lead to
$
\tilde\psi=\bigO(1) -\frac{\gamma^2}{2}\|L_{11}^{-1}\pl_1\|^2 -d_1\log\gamma
$.
An examination of  Proposition~\ref{mills refined} when $\sJ\not=\emptyset$ thus shows  that $\exp(\tilde\psi)=\bigO(\ell(\gamma))$ as $\gamma\uparrow\infty$. In other words, $\hat\ell$ is a bounded relative error estimator for $\ell(\gamma)$:
\[
\textstyle
\frac{\Var_{\bmu^*}(\hat\ell)}{\ell^2}\leq \frac{\exp(\psi(\bx^*;\bmu^*))}{\ell(\gamma)}-1\leq \frac{\exp(\psi(\breve\bx;\breve\bmu))}{\ell(\gamma)}-1\simeq \frac{\exp(\psi(\tilde\bx;\tilde\bmu))}{\ell(\gamma)}-1 =\bigO(1)\;.
\]


\subsection{Proof of Theorem~\ref{VRE}}
 In the following proof we   use the following multidimensional   Mill's ratio   \citep{savage1962mills}:    
\begin{equation}
\label{Mills}\textstyle
\frac{\Pm(A\bZ>\gamma \Sigma \bl^*)}{\phi(\gamma \Sigma \bl^*;\mathbf{0},\Sigma)}\simeq\exp\left(-\sum_k \log(\gamma l_k^*)\right),\quad\gamma\uparrow\infty\;.
\end{equation}
 This is a  generalization of the  well-known one-dimensional result:
$
\frac{\overline\Phi(\gamma)}{\phi(\gamma;0,1)}\simeq \frac{1}{\gamma},\; \gamma\uparrow\infty\;.
$
As  in the proof of Theorem~\ref{BRE}, we proceed to find  the
 asymptotic solution of the nonlinear optimization  program \eqref{optim} by considering the necessary and sufficient Karusch-Kuhn-Tucker conditions \eqref{KKT}. In the  setup of Theorem~\ref{VRE} these conditions simplify to (replacing $\bl$ with $\gamma\Sigma\bl^*$):
\begin{equation}
\label{kkt3}
\begin{split}
\bmu-\bx+\bPsi&=\mathbf{0}\\
-\bmu+(\breve L^\top-I)\bPsi+\breve L ^\top\bfeta&=\mathbf{0}\\
\bfeta\geq \mathbf{0},\;\;\gamma LL^\top\bl^*-L\bx&\leq\mathbf{0}\\
\bfeta^\top(\gamma LL^\top\bl^*-L\bx)&=0
\end{split}
\end{equation}
We can thus verify via direct substitution that the following  
\begin{equation}
\label{asymptotic soln}
\tilde\bx= \gamma L^\top\bl^*,\qquad
\tilde\bmu=\gamma (L^\top-D)\bl^*,\qquad \tilde\bfeta=o(\mathbf{1})
\end{equation}
satisfy the equations \eqref{kkt3} asymptotically. 
Equations three and four in \eqref{kkt3} are satisfied, because
$\gamma LL^\top\bl^*-L\tilde\bx=\gamma LL^\top\bl^*-L\gamma L^\top\bl^*=\mathbf{0}$. Let us now examine equations one and two in \eqref{kkt3}. First, note that from \eqref{asymptotic soln}
\[
\tilde\bl-\tilde\bmu=\gamma \breve L L^\top \bl^*-(\breve L-I)\tilde\bx-\tilde\bmu=\gamma D \bl^*>\mathbf{0}
\]
and hence from the one-dimensional Mill's ratio we have
\[
\Psi_k=\frac{\phi(\tilde l_k-\tilde\mu_k;0,1)}{\overline\Phi(\tilde l_k-\tilde\mu_k)}=\frac{\phi(\gamma D_{kk}l^*_k;0,1)}{\overline\Phi(\gamma D_{kk}l^*_k)}\simeq \gamma D_{kk}l^*_k,\qquad \gamma\uparrow\infty\;.
\]
In other words, $\bPsi\simeq \gamma D \bl^*$ as $\gamma\uparrow\infty$. It follows that for equation one  in 
\eqref{kkt3} we obtain
\[
\tilde\bmu-\tilde\bx+\bPsi=-\gamma D\bl^*+\bPsi=o(\mathbf{1})
\]
and for equation two (recall that $\breve L=D^{-1}L$, so that $\breve L^\top=L^\top D^{-1}$)
\begin{equation*}
\begin{split}
-\tilde\bmu+(\breve L^\top-I)\bPsi+\breve L ^\top\tilde\bfeta&=-\gamma(\breve L^\top-I) D\bl^*+(\breve L^\top-I)\bPsi+\breve L ^\top\tilde\bfeta\\
&=(\breve L^\top-I)(\bPsi-\gamma D\bl^*)+o(\mathbf{1})=o(\mathbf{1})\;.
\end{split}
\end{equation*}
Thus, all of the equations in \eqref{kkt3} are satisfied asymptotically and since  \eqref{kkt3} has a unique solution, we can  conclude  that $(\bx^*,\bmu^*)\simeq (\tilde\bx,\tilde\bmu)$. We now proceed to
 substitute this pair $(\tilde\bx,\tilde\bmu)$ into 
$\psi(\bx;\bmu)=\frac{\|\bmu\|^2}{2}-\bx^\top\bmu +\sum_k \log\overline\Phi(\tilde l_k-\mu_k)$. Using the one-dimensional Mill's  ratio, $\log\overline\Phi(\gamma)\simeq -\frac{1}{2}\gamma^2-\log\gamma-\frac{1}{2}\log(2\pi)$, we obtain 
\[
\sum_k\log\overline\Phi(\gamma D_{kk}l_k^*)\simeq -\frac{\gamma^2}{2}\|D\bl^*\|^2-\sum_k\log(\gamma D_{kk} l_k^*)-\frac{d}{2}\log(2\pi),\quad \gamma\uparrow\infty\;.
\]
As a consequence, using the fact that $\log|\det(L)|=\sum_k \log D_{kk}$ (recall that $L$ is triangular with positive diagonal elements),  we have
\begin{align*}
\tilde\psi=\psi( \tilde\bx;\tilde\bmu)&=\psi( \gamma L^\top \bl^*;\gamma(L^\top-D)\bl^*)\\
&= -\frac{1}{2}\|\tilde\bx\|^2+\frac{\gamma^2}{2}\|D\bl^*\|^2+\sum_k\log\overline\Phi(\gamma D_{kk}l_k^*)\\
&\simeq -\frac{\gamma^2}{2}(\bl^*)^\top LL^\top \bl^* -\frac{d}{2}\log(2\pi)-\log |\det(L)|-\sum_k\log(\gamma l_k^*)
\end{align*}
In other words, 
\[
\textstyle
\exp(\psi( \tilde\bx;\tilde\bmu))\simeq \phi(\gamma \Sigma \bl^*;\mathbf{0},\Sigma) \exp\left(-\sum_k \log(\gamma l_k^*)\right),\quad \gamma\uparrow\infty
\]
However,  by Mill's ratio \eqref{Mills}, we also have
 \[\textstyle
\Pm(A\bZ\geq \gamma\Sigma\bl^*)\simeq \phi(\gamma \Sigma \bl^*;\mathbf{0},\Sigma) \exp\left(-\sum_k \log(\gamma l_k^*)\right),\quad \gamma\uparrow\infty
\]
It follows that $\exp(\psi(\tilde \bx;\tilde\bmu))\simeq \ell(\gamma)$ and  the minimax  estimator \eqref{tilted est} exhibits vanishing relative error:
\begin{equation*}
\begin{split}
\frac{\Var_{\bmu^*}(\hat\ell)}{\ell^2}= \frac{\Em_{\bmu^*} \exp(2\psi(\bX;\bmu^*))}{\ell^2}-1
&\leq \frac{\exp\left(\psi(\bx^*;\bmu^*)\right)}{\ell}-1\\
&\simeq \frac{\exp\left(\tilde\bx;\tilde\bmu)\right)}{\ell(\gamma)}-1 =o(1),\qquad \gamma\uparrow\infty\;.
\end{split}
\end{equation*}

In contrast, for the SOV estimator $\mathring{\ell}$ we have at most bounded relative error under quite stringent conditions.
 First, the second moment on the SOV estimator satisfies
\[
\liminf_{\gamma\uparrow\infty}\Em_\mathbf{0} \exp\left(2\psi(\bX;\mathbf{0})\right) 
\geq \Em_\mathbf{0}\liminf_{\gamma\uparrow\infty} \exp\left(2\psi(\bX;\mathbf{0})\right) 
\]
 and in considering the asymptotics of $\psi(\bx;\mathbf{0})$
 we are free to select $\bx$ to obtain the best error behavior subject to the constraint  $\breve{L}\bx\geq\gamma\breve{L}L^\top\bl^*$. This gives
\[\textstyle
\exp\left(2\psi(\bx;\mathbf{0} )\right)\simeq \exp\left(2\psi(\gamma L^\top\bl^*;\mathbf{0} )\right)\simeq\textstyle \frac{1}{\gamma^{2\tr(\Lambda)}}\exp\left( -\gamma^2 (\bl^*)^\top L\Lambda L^\top\bl^* -2c_1\right),
\]
where $\Lambda=\diag([e_1,\ldots,e_d])$ is a diagonal matrix such that $e_i=\I\{\sum_jL_{ji}l_j^*>0\}$ and $c_1=\frac{\tr(\Lambda)}{2}\log(2\pi)+\sum_{k: e_k=1} \log(\sum_jL_{jk}l_j^*)$. It follows that the  relative error of the SOV estimator behaves asymptotically as
\[\textstyle
(2\pi)^{d/2}\det(L)\gamma^{d-\tr(\Lambda)}\exp\left( \frac{1}{2}\gamma^2 (\bl^*)^\top L(I-\Lambda )L^\top\bl^* -c_1+\sum_k\log l_k^*\right)\;. 
\]
$\hfill\Box$

\subsection{Proof of Corollary~\ref{corollary}}
The corollary follows from a Pinsker-type inequality \cite[Page 222, Theorem 2]{devroye1985nonparametric} by observing that (the expectation operator $\Em$ corresponds to the measure $\Pm$):

\begin{align*}
\sup_{\mathscr{A}}|\Pm(\bZ\in \mathscr{A})-\Pm_{\bmu^*}(\bZ\in \mathscr{A})|&=\frac{1}{2}\int |f(\bz)-g(\bz;\bmu^*)|\di\bz\\
&\leq \textstyle\sqrt{1-\exp\bigg(-\Em\log\frac{f(\bZ)}{g(\bZ;\bmu^*)} \bigg)}\\
&\leq \textstyle\sqrt{1-\ell(\gamma)\;\exp\left(-\psi(\bx^*;\bmu^*)\right)}\\
&\simeq \sqrt{1-\ell(\gamma)\;\exp\left(-\tilde\psi\right)}=o(1)\;,\end{align*}
where the last equality follows from
$\exp(\tilde\psi)\simeq \ell(\gamma)$, which is the case when \eqref{tilted est} is a VRE estimator. $\hfill\Box$

\subsection{Proof of Lemma~\ref{lower}}
 That $\ell_\mathrm{L}$ is a variational lower bound follows immediately from Jensen's inequality:
\begin{equation}
\label{LHS}\textstyle
\frac{1}{(2\pi)^{d/2}\sqrt{|\Sigma|}} \exp\left(- \frac{1}{2}\tr(\Sigma^{-1} \underline\Var(\bX))-\frac{1}{2} \underline\Em[\bX]^\top \Sigma^{-1} \underline\Em[\bX]-\underline\Em[\log \underline\phi(\bX)]\right)= \exp\left(\underline\Em \log \frac{ \phi(\bX;\mathbf{0},\Sigma)}{\underline\phi(\bX)}\right).
\end{equation}
Note that if $\alpha_i\idef (\ell_i-\nu_i)/\sigma_i,\;\beta_i\idef  (u_i-\nu_i)/\sigma_i,\;p_i=\overline\Phi(\alpha_i)-\overline\Phi(\beta_i)$ and $\phi(\cdot)\equiv \phi(\cdot\,;0,1)$, then all the quantities on the left-hand side are available analytically:
\begin{equation}
\label{lower eq}
\begin{split}
\textstyle
\underline\Em[X_i]&=\textstyle \nu_i+\sigma_i \frac{\phi(\alpha_i)-\phi(\beta_i)}{p_i}\\ 
\textstyle
\tr(\Sigma^{-1} \underline\Var(\bX))&=\textstyle\sum_{i=1}^d\{\Sigma^{-1}\}_{i,i}\,\sigma_i^2\Big(1+ \frac{\alpha_i\phi(\alpha_i)-\beta_i\phi(\beta_i)}{p_i} -\left(\frac{\phi(\alpha_i)-\phi(\beta_i)}{p_i}\right)^2\Big)\\
\textstyle
-\underline\Em[\log \underline\phi(\bX)]&=\textstyle\sum_{i=1}^d\frac{\alpha_i\phi(\alpha_i)-\beta_i\phi(\beta_i)}{2p_i}+\log\big(\sqrt{2\pi \exp(1)}\;\sigma_i p_i\big)
\end{split}
\end{equation}

Next, we establish the asymptotic behavior of $\ell_\mathrm{L}(\gamma)$ under the conditions of Theorem~\ref{VRE}.
Suppose the pair $(\tilde\bnu,\tilde{\boldsymbol\sigma})$ satisfies $\mathrm{diag}^2(\tilde{\boldsymbol\sigma})\simeq\Sigma$   and $\tilde\bnu\simeq \bl-\gamma\mathrm{diag}^2(\tilde{\boldsymbol \sigma})\bl^*=\gamma(\Sigma-\mathrm{diag}^2(\tilde{\boldsymbol \sigma}))\bl^*$ as $\gamma\uparrow \infty$. Then,  $\boldsymbol\alpha\simeq \gamma \mathrm{diag}(\tilde{\boldsymbol \sigma})\bl^*$, which in combination with
 $\log\overline\Phi(\gamma)\simeq -\frac{1}{2}\gamma^2-\log(\gamma)-\frac{1}{2}\log(2\pi)$,  implies
 $\underline\Em[\bX]\simeq \gamma \Sigma\bl^*  $.
Hence, substituting  $(\tilde\bnu,\tilde{\boldsymbol\sigma})$ into \eqref{lower eq} and then into the left-hand-side of \eqref{LHS}, and  simplifying, we obtain
\begin{align*}
\ell(\gamma)\geq \ell_\mathrm{L} &\geq \textstyle \frac{1}{(2\pi)^{d/2}\sqrt{|\Sigma|}} 
\exp\left(-\frac{1}{2} \underline\Em[\bX]^\top \Sigma^{-1} \underline\Em[\bX]+\frac{1}{2}\sum_i \left(\frac{\phi(\alpha_i)}{\overline\Phi(\alpha_i)}\right)^2+\sum_i \log(\sqrt{2\pi}\;\tilde\sigma_i \overline\Phi(\alpha_i)) \right)\\
&\simeq \textstyle \frac{1}{(2\pi)^{d/2}\sqrt{|\Sigma|}} \exp\left(-\frac{1}{2} (\gamma\Sigma\bl^*)^\top\Sigma^{-1}(\gamma\Sigma\bl^*)-\sum_{i}\log(\alpha_i/\tilde\sigma_i)\right)\\
&\simeq\textstyle \frac{1}{(2\pi)^{d/2}\sqrt{|\Sigma|}} 
\exp\left(-\frac{\gamma^2}{2} (\bl^*)^\top\Sigma\bl^*-\sum_{i}\log(\gamma l_i^*)\right)\simeq \ell,\qquad \gamma\uparrow\infty
\end{align*}
where the last asymptotic equivalence follows from \eqref{Mills}.   
Finally, the convergence of \eqref{total var} follows by applying the Pinsker-type inequality \citep{devroye1985nonparametric} in conjunction with
$\sqrt{1-\exp\left(-\underline\Em 
\log\frac{\underline \phi(\bX)}{f(\bX)}\right)}=\sqrt{1- \frac{1}{
\ell}\exp\left(\underline\Em \log\frac{\phi(\bX;\mathbf{0},\Sigma)  }{\underline 
\phi(\bX)} \right)}\leq \sqrt{1-\ell_{\mathrm{L}}/\ell}= o(1)$. $\hfill\Box$

\bibliographystyle{chicago}
\bibliography{normcdf}
\end{document}